\documentclass[a4paper,useAMS,usenatbib]{mnras}

\usepackage{graphicx}
\usepackage{amsmath}

\title[Galaxy Evolution Paradigm from
Herschel]{The New Galaxy Evolution Paradigm Revealed by the Herschel Surveys}
\author[Stephen Eales]{Stephen Eales$^{1}$\thanks{E-mail:
sae@astro.cf.ac.uk}, Dan Smith$^2$, Nathan Bourne$^3$, Jon Loveday$^4$, Kate Rowlands$^5$, 
\newauthor
Paul van der Werf$^6$,
Simon Driver$^7$, Loretta Dunne$^{1,3}$, Simon Dye$^8$, 
\newauthor
Cristina Furlanetto$^{8}$, 
R.J. Ivison$^{9,3}$, 
Steve Maddox$^{1,3}$, Aaron Robotham$^7$,
\newauthor
Matthew W.L. Smith$^1$, 
Edward N. Taylor$^{10}$,
Elisabetta Valiante$^1$, Angus Wright$^{7,11}$,\newauthor 
Philip Cigan$^1$, 
Gianfranco De Zotti$^{12,13}$, 
Matt J. Jarvis$^{14,15}$, 
Lucia Marchetti$^{16}$, \newauthor
Micha{\l} J.~Micha{\l}owski$^{17}$,
Steven Phillipps$^{18}$, Sebastien Viaene$^{19}$\newauthor
and Catherine 
Vlahakis$^{20}$\\ 
$^{1}$School of Physics and Astronomy, Cardiff University, The Parade, Cardiff CF24 3AA, UK\\
$^2$Centre for Astrophysics Research, 
School of Physics, Astronomy and Mathematics, University of Hertfordshire,\\
College Lane, Hatfield, AL10 9AB, UK\\
$^3$Institute for Astronomy, The University of Edinburgh, Royal Observatory, Blackford Hill,
Edinburgh, EH9 3HJ, UK\\
$^4$Astronomy Centre, University of Sussex, Falmer, Brighton BN1 9QH, UK\\
$^5$Department of Physics and Astronomy, JHU, Bloomberg Center, 3400
N. Charles St., Baltimore, MD 21218, USA\\
$^6$ Leiden Observatory, PO Box 9513, 2300 RA Leiden, the Netherlands\\
$^7$International Centre for Radio Astronomy Research, 7 Fairway, The University of Western\\
Australia, Crawley, Perth, WA 6009, Australia\\
$^8$School of Physics and Astronomy, University of Nottingham, University Park, Nottingham
NG7 2RD, UK\\
$^9$European Southern Observatory, Karl-Schwarzschild-Strasse 2, 85748, Garching, Germany\\
$^{10}$ Centre for Astrophysics and Supercomputing, Swinburne University
of Technology, Hawthorn 3122, Australia\\
$^{11}$ Argelander-Institut fur Astronomie, Auf dem Hugel 71, D-53121 Bonn,
Germany\\
$^{12}$ INAF-Osservatorio Astronomico di Padova, Vicolo Osservatorio 5, I-35122 Padova, Italy\\
$^{13}$ SISSA, Via Bonomea 265, I-34136 Trieste, Italy\\
$^{14}$ Astrophysics, Department of Physics, Keble Road, Oxford, OX1 3RH, UK\\
$^{15}$ Physics and Astronomy Department, University of the
Western Cape, Bellville 7535, South Africa\\
$^{16}$ Department of Physical Sciences, The Open University, Milton Keynes,
MK7 6AA, UK\\
$^{17}$ Astronomical Observatory Institute, Faculty of Physics,
Adam Mickiewicz University, ul.~S{\l}oneczna 36, 60-286 Pozna{\'n}, Poland\\
$^{18}$ Astrophysics Group, Department of Physics, University of Bristol, Tyndall Avenue,
Bristol BS8 1TL\\
$^{19}$ Sterrenkundig Observatorium,Universiteit Gent, Krijgslaan 281 S9, B-9000 Gent, Belgium\\
$^{20}$ North American ALMA Science Center, National Radio Astronomy Observatory,
520 Edgemont Road, Charlottesville, VA 22901, USA}

\begin{document}

\pagerange{\pageref{firstpage}--\pageref{lastpage}} \pubyear{2002}

\maketitle

\label{firstpage}

\newpage

\begin{abstract}

The {\it Herschel Space Observatory} has
revealed a very different galaxyscape from
that shown by optical surveys which presents
a challenge for galaxy-evolution models.
The {\it Herschel} surveys reveal (1) 
that
there was rapid galaxy evolution in the very recent past and (2) that
galaxies
lie on a
a single Galaxy Sequence (GS) rather than 
a star-forming `main sequence' and a separate region of `passive' or `red-and-dead' galaxies.
The form of the GS is now clearer because
far-infrared
surveys such as the Herschel ATLAS pick up a population of
optically-red star-forming galaxies that would have been classified as 
passive using most optical criteria. The space-density of this population
is at least as high
as the traditional star-forming population.
By stacking
spectra of H-ATLAS galaxies over the redshift range $0.001 < z < 0.4$, we show
that the galaxies responsible
for the rapid low-redshift evolution
have high stellar masses, high star-formation rates but,
even several billion years in the past, old stellar populations---they are thus likely to be
relatively recent ancestors of early-type galaxies in the Universe today.
The form of the GS is inconsistent with rapid quenching models
and
neither the analytic bathtub model nor the hydrodynamical
EAGLE simulation can reproduce the rapid cosmic evolution.
We propose a new gentler model of galaxy evolution
that can explain the new {\it Herschel} results and
other key properties of the galaxy population. 

\end{abstract}


\section{Introduction}

Over the last decade a simple phenomenological
model of galaxy evolution has been become widely used
by astronomers to interpret
observations. In this model, 
star-forming galaxies lie on
the `Galaxy Main Sequence' (henceforth GMS), a distinct
region in a plot of star-formation rate versus galaxy stellar
mass
(e.g Noeske et al. 2007; Daddi et al. 2007; Elbaz et al. 2007;
Peng et al. 2010; Rodighiero
et al. 2011; Whitaker et al. 2012; Lee et al. 2015).
Over cosmic time, the GMS gradually moves downwards in star-formation rate,
which decreases
by a factor of $\simeq$20 from a redshift of 2 to the current epoch
(Daddi et al. 2007). Observations of the molecular gas and dust in galaxies 
show that the principal cause of this evolution is that high-redshift galaxies
contained more gas and therefore formed stars at a faster rate (Tacconi et
al. 2010; Dunne et al. 2011; Genzel et al.
2015; Scoville et al. 2016).

In this phenomenological paradigm an individual galaxy moves along the GMS
until some process quenches
the star formation in the galaxy, which then moves rapidly (in cosmic
terms) across the diagram to the region occupied by `red and dead' or
`passive' galaxies. Possible quenching processes include galaxy merging
(Toomre 1977), with a starburst triggered by
the merger rapidly using up the available gas; the expulsion of gas by a wind from an 
active galactic nucleus
(Cicone et al. 2014); the rapid motion
of star-forming clumps towards the centre of the galaxy
(Noguchi 1999; Bournaud et al. 2007; Genzel et al.
2011, 2014), leading to the formation of a stellar bulge, which then
stabilizes the star-forming gas disk and reduces the rate at which the gas collapses
to form stars (Martig et al. 2009); and a variety of environmental processes which either reduce the rate
at which gas is supplied to a galaxy or which drive out most of the existing gas 
(Boselli and Gavazzi 2006).

As can be seen from the long list of possible quenching mechanisms,
the physics underlying this paradigm is unknown.
Peng et al. (2010) have shown that many statistical
properties of star-forming and passive galaxies can be explained by a model
in which both the star-formation rate 
and the probability of quenching are proportional to the galaxy's stellar
mass, but the physics behind both
proportionalities is unknown.
Although it is clear that the increased star-formation rates
in high-redshift galaxies are largely due to their increased gas
content, there is also evidence that the star-formation efficiency
is increasing with redshift (Rowlands et al. 2014; Santini
et al. 2014; Genzel et al. 2015; Scoville et al. 2016); so
either the physics of star formation or the properties of
the interstellar gas (Papadopoulos and Geach 2012) must
also be changing with redshift in some way.

Implicit in this paradigm is the assumption that there
are two physically-distinct 
classes of galaxy. These two classes of galaxy are variously called `star-forming' and `passive',
`star-forming' and `red-and-dead' or `star-forming' and `quenched'. 
The most visually striking evidence that there are two separate classes
is that
on plots of optical colour versus optical absolute magnitude,
galaxies fall in two distinct areas: a `blue cloud' of star-forming
galaxies and a tight `red sequence' representing the passive galaxies
(e.g. Bell et al. 2004). However, in an earlier paper
(Eales et al. 2017) we argued that the the tight red sequence
is the result of optical colour depending
only very weakly on specific star-formation rate (star-formation rate
divided by galaxy stellar mass, henceforth SSFR) for $\rm SSFR < 5 \times 10^{-12}\ year^{-1}$;
the red sequence is therefore better thought of
as the accumulated number of galaxies that have passed below this
critical SSFR, all of which 
pile up at the same colour, rather than representing a distinct class of galaxy.
The two
classes are largely the same as the morphological classes of early-type 
and late-type galaxies
(henceforth ETGs and LTGs).
Although there is plenty of evidence for a gradual change
in the properties of galaxies along the morphological Hubble sequence
(e.g. Kennicutt 1998), there is now little
evidence
for a clear dichotomy between the two broad morphological classes (Section 5.5 of this paper).

The launch in 2009 of the {\it Herschel Space Observatory} (Pilbratt et al. 2010) 
gave a different view of the galaxy population
from the one
given by optical surveys. Apart from the interest of this new
galaxyscape, produced by the different selection effects
operating on optical and far-infrared surveys (\S 2), 
{\it Herschel's} launch provided two immediate practical benefits for
astronomers studying galaxy evolution.
The first was that {\it Herschel} made it possible
for astronomers to directly
measure the part of the energy output of stars that is hidden by dust. 
For example, by using {\it Herschel} and other telescopes
to measure the 
bolometric luminosity
of different galaxy classes, it is possible to measure the size
of the morphological transformation that has occurred in the
galaxy population in the last eight billion years (Eales
et al. 2015). 
The second benefit is
that with {\it Herschel} submillimetre photometry, which
now exists for $\sim10^6$ galaxies, it is possible to estimate
the mass of a galaxy's ISM from its dust emission
(Eales et al. 2012; Scoville et al. 2014; Groves et al. 2015), a technique
that has since been profitably extended to ALMA observations (Scoville et al.
2016).

In this paper, we investigate this new galaxyscape.
The paper is based on two {\it Herschel} surveys:
the {\it Herschel} Reference Survey (henceforth HRS) and
the 
{\it Herschel} Astrophysical Terrahertz Large Area Survey (henceforth
H-ATLAS). We describe these surveys in more detail in Section 2
but, briefly, the HRS is a volume-limited sample of 323 galaxies,
designed before launch to be a complete as possible census of the stellar mass
in the Universe today; each galaxy was then individually observed with
{\it Herschel} (Boselli et al. 2010; Smith et al. 2012b; 
Cortese et al. 2012; Ciesla
et al. 2012;
Eales et al. 2017).
The H-ATLAS was the largest (in sky area, 660 square degrees) 
{\it Herschel} extragalactic survey, consisting
of imaging
at 100, 160, 250, 350 and 500 $\mu$m
of five fields,
two large fields at the North and South Galactic Poles
and three smaller fields on the celestial equator (Eales et al. 2010).

There is already one important advance in our knowlege of galaxy
evolution provided by the {\it Herschel} surveys, although we suspect
this has yet been absorbed by the wider astronomical community.
In an early H-ATLAS paper (Dye et
al. 2010), we showed that the 250-$\mu$m luminosity function evolves remarkably
rapidly, even showing
significant evolution by a redshift of 0.1, 
which we have confirmed recently with a much larger dataset (Wang
et al. 2016a). We have also shown (Dunne et al. 2011; Bourne et al. 2012)
that there is rapid evolution
in the masses of dust in galaxies, and therefore
in the mass of the ISM.
Using radio continuum emission to trace the star formation,
we found that
there is also rapid evolution at low redshift in the star-formation rates
of galaxies (Hardcastle et al.
2016). Marchetti et al. (2016) reached the same conclusion 
from a different {\it Herschel} survey and using
a different method of estimating the star-formation rate
(from the bolometric dust luminosity).
This rapid low-redshift evolution is important because, as we show in this paper,
it is not predicted by important galaxy-evolution models.

A note on nomenclature: in this paper, we generally
use the term `Galaxy Sequence' rather than `Galaxy Main Sequence'. 
The latter term is implicitly based on the phenomenological paradigm, in which
galaxies, like stars, 
spend most of their
life in an active phase, which then comes to a definite end. We prefer the empirical
term `Galaxy Sequence', which is free of theoretical assumptions. Our definition of the
term is that it refers to 
the distribution of galaxies in a plot of 
specific star-formation rate versus stellar mass that contains
most of the stellar mass in a given volume of space.

The arrangement of this paper is as follows.
In Section 2 we describe the two {\it Herschel} surveys in
more detail.
Section 3 and 4 describes results from the two
surveys that have implications for galaxy evolution.
In Section 5, we discuss the implications of these results for
the phenomenological paradigm and propose an alternative model
for galaxy evolution that is in better agreement with these
results. We suggest that readers not interested in the
details
of the {\it Herschel} results but interested in their implications
skip to
Section 5, which we start with a summary of the main
observational results. 
A summary of the main results of this
paper is given in Section 6.

We assume a Hubble constant of 67.3 $\rm km\ s^{-1}\ Mpc^{-1}$
and the other {\it Planck} cosmological parameters  
(Planck Collaboration 2014).

\section[]{The Herschel Surveys}

 The HRS consists of 323 galaxies with distances between 15 and 25 Mpc
and with a near-infrared K-band limit of $K < 8.7$ for early-type
galaxies (ETGs) and $K < 12$ for late-type galaxies (LTGs, Boselli et al. 2010).
The sample was designed to be a volume-limited sample of galaxies
selected on the basis of stellar mass. Eales et al. (2017) estimate
that within the HRS volume the survey is complete for LTGs with stellar
masses above $\simeq 8 \times 10^8\ M_{\odot}$ and for ETGs with
stellar masses above $\simeq  2 \times 10^{10}\ M_{\odot}$. The survey
therefore misses low-mass ETGs, but Eales et al. (2017) show that there
is very little mass contained in these objects: $\simeq$90\% of the
total stellar mass in ETGs with masses $>10^8\ M_{\odot}$ 
in the HRS volume is contained in the galaxies in the sample.
As a result of the 
{\it Herschel} photometry (Ciesla et
al. 2012; Smith et al. 2012; Cortese et al.
2014) and the proximity of the galaxies, 
there are extremely sensitive measurements of the dust continuum emission in
five far-infrared bands for each of the HRS galaxies.
Even though ETGs are often assumed to contain very little dust, Smith
et al. (2012b) detected continuum dust emission from
31 of the 62 HRS ETGs and obtained very tight
limits on the amount of dust in the remainder.

High-quality photometry in 21 photometric bands, from the $ UV$ to
the far-infared, makes the HRS ideal for the application of galaxy modelling
programs such as MAGPHYS (Da Cunha et al. 2008). 
De Vis et al. (2017) applied MAGPHYS to the HRS galaxies, obtaining
estimates of key galaxy properties such as star-formation rate and stellar
mass. Eales et al. (2017) used
these results to look at the relationship between specific star-formation rate
and stellar mass in the HRS
volume (Figure 1), finding that
galaxies follow a smooth curved Galaxy Sequence (GS),
with a gradual change in galaxy morphology along the sequence
and no abrupt change between LTGs and ETGs. 
They showed that the location and shape of the GS in Figure 1 is consistent
with other recent attempts either to
plot the entire GS (Gavazzi et al. 2016)
or to plot the part of the GS classified as star-forming (Renzini and Peng 2015).
Oemler et al. (2017, O17) have recently carried out a reanalysis of the
SDSS galaxy survey, taking account of several selection effects, and have
found a galaxy distribution very like that in Figure 1.
Since Figure 1 contains all the LTGs in the HRS volume with
masses $\succeq 8 \times 10^8\ M_{\odot}$, and since
there is very little
stellar mass in the ETGs in its bottom-left-hand corner, the diagram should be
a good
representation of where the stars in the Universe are today,
after 12 billion years of galaxy evolution.

\begin{figure}
\includegraphics[trim=0mm 2mm 0mm 0mm, clip=True, width=0.49\textwidth]{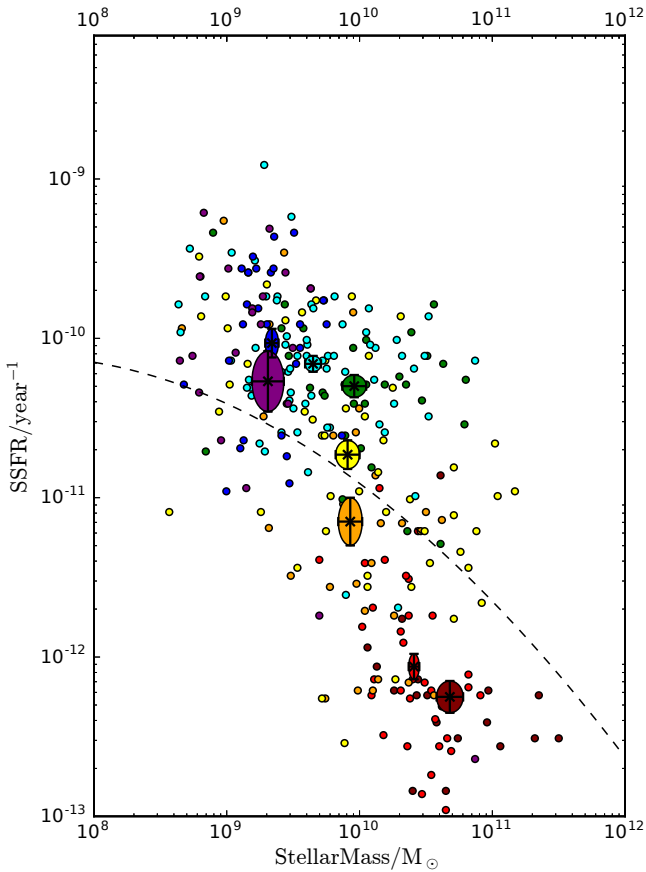}
  \caption{specific star-formation rate versus stellar mass for
the galaxies in the {\it Herschel} Reference Survey, a volume-limited
sample designed to contain most of the stellar mass in the survey
volume, reproduced from Eales et al. (2017 - see that paper
for more details).
The colours show the morphologies of the
galaxies:
maroon - E and E/S0; red - S0;
orange - S0a and Sa; yellow - Sab and Sb; green - Sbc;
cyan - Sc and Scd; blue - Sd, Sdm; purple - I, I0, Sm and Im.
The coloured ellipses show the $1\sigma$ error region on the mean
position for each morphological class, with the colours
being the same as for the individual galaxies.
The dashed line shows the results of fitting a second-order polynomial
to the 
H-ATLAS galaxies (Section 3), using a method that corrects
for the effect of Malmquist bias.
Note the consistency in where the GS lies, whether its location
is obtained from a volume-limited survey such as the HRS
or a far-infrared survey such as H-ATLAS.
}
\end{figure}

While the HRS was a sample of galaxies selected in the near-infrared,
which were then observed in the far-infared with {\it Herschel}, the H-ATLAS
was a `blind' far-infrared survey in which the galaxies were
selected based on their far-infrared flux density.
In its five fields the H-ATLAS detected $\sim500,000$ sources.
The fields we use in this paper are the three small fields
on the celestial equator, which cover a total area
of 161.6 square degrees and were the same fields surveyed
in the Galaxy and Mass Assembly project (henceforth GAMA).
GAMA was a deep spectroscopic survey (Driver et al.
2009; Liske et al. 2015) complemented with matched-aperture
photometry
throughout the $UV$, optical and IR wavebands (Driver et al.
2016). We used the 
optical SDSS images
to find the galaxies producing the Herschel sources and then the
GAMA data to provide redshifts and
matched-aperture photometry for these galaxies.

We have recently released our final images and catalogues for the GAMA fields
(Valiante at al. 2016; Bourne et al. 2016\footnote{This dataset
can be obtained from h-atlas.org}). The catalogue of {\it Herschel} sources
(Valiante et al. 2016)
contains 120,230 sources detected at $>4\sigma$ at 250, 350 or 500 $\mu$m,
of which 113,995 were detected above this signal-to-noise
at 250 $\mu$m, our most
sensitive wavelength, which corresponds to a flux-density limit 
of $\simeq$30 mJy.
We have also released a catalogue of 44,835 galaxies which are the probable sites of
the {\it Herschel} sources (Bourne et al. 2016).
We found these galaxies by looking for galaxies on the
the r-band SDSS images close to the positions of the {\it Herschel} sources;
we then used
the magnitude of the galaxy and its distance from the
{\it Herschel} source to estimate
a Bayesian probability
(the `reliability' in our nomenclature) that the galaxy is
producing
the far-infrared emission.

Our base sample in this paper are the 19556 galaxies in this catalogue
detected at $>4 \sigma$ at 250 $\mu$m, with matched-aperture multi-wavelength
photometry and 
spectroscopic redshifts
in the redshift range $0.001 < z < 0.4$, the lower redshift limit chosen
to minimize the effect of galaxy peculiar motions.
The redshift distribution
of this sample is shown in Table 1.

There are several possible sources of error that we need to consider. The first is
the possibility that we have incorrectly associated a {\it Herschel} source and
an SDSS galaxy. We can estimate the number of sources that may have been
misidentified in this way by adding up $1-R$ for all the sources, where $R$ is
the reliability. We calculate that of our base sample, 435 sources (2.2\%) have been
incorrectly associated with SDSS galaxies. 

The second is that there are an additional
819 galaxies which satisfy the other criteria above but which
do not have multi-wavelength aperture-matched photometry, since their
redshifts were measured after the completion of the
GAMA photometry program. These are
essentially random omissions from the sample and are thus very unlikely to
have any effect on the results in sections 3 and 4 (these 
objects are included in the investigation of the evolution of the
luminosity function in Section 4.3).

A more important issue is the possibility that we have missed associations.
Bourne et al. (2016) have estimated, as a function of source redshift, the
probability that we will have found the galaxy producing the
{\it Herschel} source. Their estimates are shown in Table 1, which range from
91.3\% in the redshift range $0.001 < z < 0.1$ to 72.2\% in the highest redshift
bin, $0.3 < z < 0.4$. 

The final source of error is the requirement
that the galaxy has a spectroscopic redshift, which we have made so as
not to introduce any additional errors into our spectral fits (Section
3). We can estimate the overall completeness of the base sample by
finding the additional galaxies that have photometric redshifts in
the redshift range $0.001 < z < 0.4$ but which do not have spectroscopic
redshifts. There are 5701 of these galaxies, which implies the spectroscopic sample
is 78\% complete. However, we expect the completeness
of the base sample to be a strong function of redshift. We investigated how
the completeness varies with redshift using the following method. The magnitude
limit of the GAMA spectroscopic survey was $r=19.8$ (Liske et al.
2015). We can guage the possibility that we have missed galaxies
because they are fainter than this limit by counting the number
of galaxies in the base sample that 
fall in the half magnitude {\it brighter} than the spectroscopic limit:
$19.3 < r < 19.8$. If this number is small we would not expect
incompleteness from this effect to be an issue. We list these percentages
in Table 1. In the lowest redshift bin, the percentage is very small
(0.7\%), but it is very high in the two highest redshift bins. Therefore,
the lowest-redshift bin should not be significantly
affected but the two highest-redshift bins will be significantly
incomplete. 
The incompleteness
will be most severe for galaxies with low stellar masses.

In summary, there are number of sources of systematic error associated with the
method used to find the galaxies producing the
{\it Herschel} sources. The numbers in Table 1 show that
these errors are likely to be quite small for
the lowest-redshift bin but large for the two highest-redshift bins,
making the base sample highly incomplete at $z > 0.2$.

\begin{table*}
\caption{The H-ATLAS Base Sample}
\begin{tabular}{cccccc}
\hline
Redshift & No. &  ID fraction & Last & bad & GAMA \\
range & & & 0.5 mag & fits & \\
\hline
$0.001 < z < 0.1$ & 3,456 & 91.3\% & 0.7\% & 12.1\% & 17,768 \\
$0.1 < z < 0.2$ &  7,096 &  87.7\% & 5.1\% & 7.1\% & 38,768 \\
$0.2 < z < 0.3$ & 5,400 &  80.4\% & 22.3\% & 9.2\% & 21,323 \\
$0.3 < z < 0.4$ & 3,604 & 72.2\% & 39.5\% & 12.8\% & 7,289 \\
\hline
\end{tabular}

\medskip
\begin{flushleft}
Notes: Col. 1 - redshift range; col. 2 - No. of galaxies
in base sample - these are the galaxies used in the anlysis in this paper; 
col. 3 - Estimated percentage of
H-ATLAS sources in this redshift range for which our search procedure
should have found the galaxies producing the submillimetre emission
(Bourne et al.
2016); col. 4 - percentage
of galaxies in column 2 with r-band magnitude
in the range $19.3 < r < 19.8$ (see text for
significance);
col. 5 -  Percentage that were excluded from the base sample
because there was $<$1\% probability that the best-fit
MAGPHYS SED was a good fit to the
multi-wavelength photometry; col. 6 - No. of galaxies
from the GAMA survey in this redshift range (Driver et al. 2009; Liske
et al. 2015).
\end{flushleft}
\end{table*}

On top of these errors, there is
is the unavoidable selection effect found in
all flux-density-limited surveys: Malmquist bias.
The H-ATLAS is biased towards galaxies with high 250-$\mu$m luminosities
in the same way that an optical survey such as the SDSS is biased towards
galaxies with high optical luminosities.
Thus galaxies with low masses of interstellar dust, such as ETGs, will be under-represented
in H-ATLAS. For example, the ETGs in the HRS have a mean dust mass of $\rm
\sim 10^5\ M_{\odot}$ (Smith
et al. 2012b), but there are only 22 galaxies in our base sample
with dust-mass estimates 
(\S 3) less than $10^{5.5}\ M_{\odot}$. 
In the next section we will make an attempt to
correct H-ATLAS for the effect of Malmquist bias.

The result of selection effects
is that a very different galaxy population
is found in 
a submillimetre survey such as H-ATLAS
from an optical survey such as the SDSS.
In optical colour-versus-absolute-magnitude diagrams, galaxies
detected in optical
surveys lie in a `blue cloud' or on a `red sequence' with
a `green valley' in between. We show in an accompanying paper
(Eales et al. in preparation) that 
the distribution of H-ATLAS galaxies on the same
diagram is almost the opposite of this, with the far-infared-selected
galaxies
forming a `green mountain'. We show in the accompanying
paper that both distributions are the natural result of 
selection effects operating on the smooth GS shown in Figure 1.
In the next section, we start from our biased sample of galaxies
detected in H-ATLAS\footnote{No more biased,
of course, than an optical survey.} and investigate whether the GS we obtain after
correcting for selection effects is consistent with the
GS we see in Figure 1.

\section[]{The H-ATLAS Galaxy Sequence}

\subsection{Method}

The main purpose of the work described in this
section was to investigate whether the low-redshift
GS derived from H-ATLAS, after correcting for selection
effects, is consistent with the GS derived from the
{\it Herschel} Reference Survey that is shown in Figure 1.
As for the HRS, we used MAGPHYS 
(Da Cunha, Charlot and Elbaz 2008)
to estimate the specific star-formation rates and stellar
masses of the galaxies in the base sample, which
all have high-quality matched-aperture photometry
in 21 bands from the
ultraviolet, measured with
the 
{\it Galaxy Evolution
Explorer}, to the five {\it Herschel}
measurements in the far infrared.

For the reader that is not familiar with the model,
we give here very brief details of the model
and our application of it.
 MAGPHYS is a model of a galaxy based on the model
of the ISM of Charlot and Fall (2000), who investigated the
effects on a galaxy's spectrum and SED of the newly-formed stars
being more obscured by dust than the older stellar population, because they are
still surrounded by the dust in their natal giant molecular clouds.

MAGPHYS generates
50,000 possible models of the SED
of an unobscured stellar population, ultimately based on
the stellar synthesis models of Bruzual and Charlot (2003), and
50,000 models of the dust emission from the
interstellar medium. By linking the two sets of models
using a dust obscuration model that balances the radiation
absorbed at the shorter wavelengths with the energy emitted
in the infrared, the program generates 
templates which are then fitted to the galaxy photometry.
From the quality of the fits between the templates and the
measurements, the program produces probability distributions
for important global properties of each
galaxy.
An advantage of the model is 
that the large number of templates make
it possible to generate
a probability distribution for each galaxy property.
MAGPHYS uses the stellar initial mass function
from Chabrier (2003).

Our detailed procedure was the same,
with the exceptions 
listed below, 
as that described by Smith et al.
(2012a), who applied MAGPHYS to the galaxies
in the small H-ATLAS field observed as part
of the {\it Herschel} Science Demonstration Phase.
As in the earlier paper,
the value we use for each galaxy property
is the median value 
from the probability distribution returned by MAGPHYS,
since this is
likely to be the most robust estimate (Smith et al. 2012a).
The star-formation rate we use is the average star-formation rate
over the last 0.1 Gyr.

The biggest change from the earlier work was that we replaced the UKIRT near-infrared
photometry and IRAS photometry with the photometry in five near-infrared
bands ($z$, $Y$, $J$, $H$ and $K_s$) from
the VISTA Kilo-Degree Infrared Galaxy Survey (VIKING, Edge et al.
2013) and in the four bands measured with the
{\it Wide-Field Infrared Survey Exlorer}.
A minor change was in the calibration errors
used for photometry measured with
the two cameras on {\it Herschel}, PACS and SPIRE, which we reduced
from the values used in our earlier paper
to 5.5\% for
SPIRE and 7\% for PACS, the values recommended by Valiante et
al. (2016).
We excluded galaxies from the base sample for
which there was $<$1\% probability that the best-fit
MAGPHYS SED was a good fit
to the multi-waveband photometry. The number of these
objects is shown in Table 1. Smith et al. (2012a)
did a detailed examination of these objects and concluded
that the vast majority are
due to serious problems with the aperture-matched
photometry,
probably
due to neighbouring objects within the aperture.
We have checked that none of the results
in this paper is spuriously generated by the
exclusion of these object by repeating the analysis with
them included, obtaining similar results. 

We have assumed that the SEDs are dominated by emission that is directly or
indirectly from stars. This assumption is supported
by the results of
Marchetti et al. (2016), who, using {\it Spitzer}
data, concluded 
that only $\simeq$3\% of the galaxies at $z < 0.5$ in the HerMES
Wide sample, a {\it Herschel} sample with a similar depth to our own, have
an SED dominated by emission from an AGN.
Many of the H-ATLAS galaxies which do have an AGN-dominated SED will
anyway have been eliminated by the requirement that MAGPHYS produces a good
fit to the measured SED.

The results from MAGPHYS have been checked in a number of ways.
Eales et al. (2017) showed that 
the MAGPHYS stellar mass estimates for the HRS galaxies agree
well with the estimates of Cortese et al. (2012), who estimated
stellar masses from i-band
luminosities and a relation between mass-to-light ratio and
g-i colour from Zibetti et al. (2009). In the same way, 
the MAGPHYS estimates of stellar mass 
for the H-ATLAS galaxies agree
well with estimates from the optical spectral energy distributions (Taylor
et al. 2011). We note, however, that these comparisons have been
made with studies that are ultimately based on the stellar
synthesis models and initial mass function on which MAGPHYS is based.

We do not have independent measurements of the star-formation rate
with which to test the MAGPHYS estimates.
However, in a comparison of 
12 different methods of estimating star-formation rates,
Davies et al. (2016) showed that
the use of MAGPHYS estimates 
does not lead to any biases in the relationship between SSFR
and galaxy stellar mass (see their Fig. 10).

Finally Hayward and Smith (2015) have demonstrated, using 
simulated galaxies,
that MAGPHYS is very reliable for
estimating galaxy stellar masses and star-formation rates,
irrespective of star-formation history, viewing angle,
black hole activity etc.

\subsection{Results}

Figure 2 shows specific star-formation rate (SSFR) plotted against galaxy stellar mass for the
H-ATLAS galaxies in four redshift
bins. The errors in the estimates of the logarithm of SSFR are typically
0.2 but can be much larger for galaxies at the bottom of the diagram.

A simple argument shows one of the effects of Malmquist bias.
The H-ATLAS galaxies are selected based on their continuum
dust emission and thus the sample will be biased towards
galaxies with a large mass of dust, and thus consequently
a large ISM mass and a high star-formation rate.
Since lines of constant star-formation rate run roughly parallel
to the galaxy distributions in Figure 2, 
the absence of galaxies to the bottom left of each panel may well
be the result of Malmquist bias. Conversely, however, the upper envelope of
each distribution, and its negative gradient, should not be
significantly affected by this.

The distribution in the lowest redshift bin appears 
curved. To assess whether this is statistically significant,
we fitted both a straight line and a polynomial to the distribution, minimising
the sum of $\chi^2$ in the SSFR direction. The
polynomial had the form:

\smallskip
\begin{align}
log_{10}(SSFR) = a + & b \times (log_{10}M_* - 10.0)\nonumber\\
&  + c \times (log_{10}M_*-10.0)^2
\end{align}
\smallskip

\noindent The best-fit polynomial is shown in Figure 3.
The reduction in the total value of
$\chi^2$  obtained from using a polynomical rather than a straight line is 
336.
Since the expected reduction in $\chi^2$ when fitting a function
with one additional parameter is itself distributed as $\chi^2$ with one degree
of freedom, the reduction in $\chi^2$, and thus the
curvature in the
low-redshift GS, is highly significant.
This adds to the other recent evidence that the GS is curved, whether only star-forming
galaxies are plotted (Whitaker et al. 2014; Lee et al. 2015; Schreiber
et al. 2016; Tomczak et al. 2016) or all galaxies are plotted
(Gavazzi et al. 2015; Oemler et al. 2017 - henceforth O17).

\begin{figure}
\includegraphics[trim=0mm 10mm 0mm 0mm, clip=True, width=64mm]{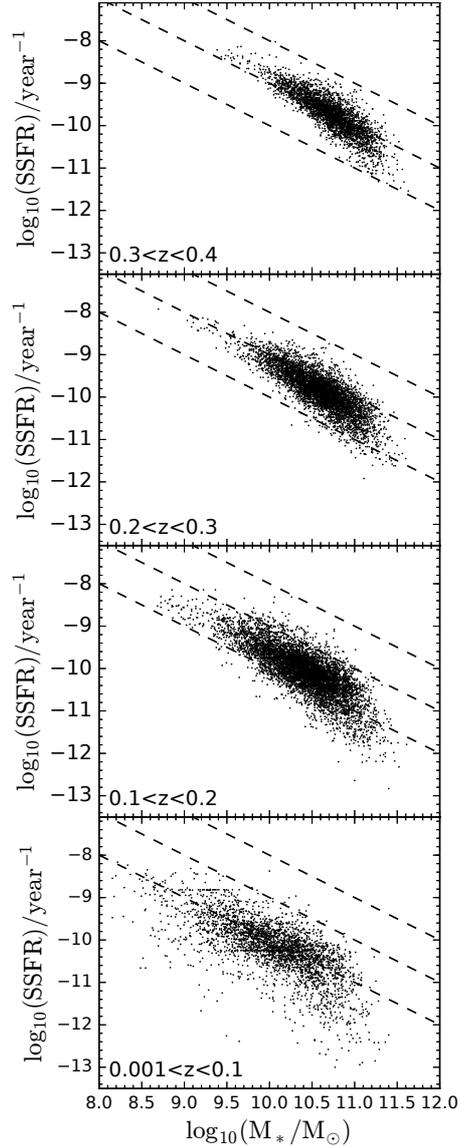}
  \caption{Specific star-formation rate versus galaxy stellar mass in four redshift bins.
The three dashed lines correspond to star-formation rates of 1, 
10 and 100 $\rm M_{\odot}\ year^{-1}$.
}
\end{figure}

We used the following method to correct for the
effect of Malmquest bias in Figure 3.
We divided the SSFR-versus-stellar-mass diagram into rectangular bins 
and calculated the following quantity in each bin:
\smallskip
\begin{align}
N(SSFR,M_*) = \sum_i {1 \over V_{acc,i}}
\end{align}
\smallskip

\noindent where the sum is over all the galaxies in that bin and $V_{acc,i}$ is the
accessible volume of the i'th galaxy, the volume in which that galaxy could still
have been detected above the 250-$\mu$m flux limit.
This is given by:

\smallskip
\begin{align}
V_{acc,i} = \int_{z_{min}}^{z_{max}} dV
\end{align}

\noindent In this equation, $z_{min}$ is the lower redshift limit (0.001)
and $z_{max}$ is the lower of (a) the upper redshift limit of the redshift bin and (b) the
redshift at which the flux density of the galaxy would equal the 250-$\mu$m flux limit
of the sample. 

This technique is the standard technique for correcting for the
effect of accessible volume. 
It will produce an unbiased estimate of
$N(SSFR,M_*)$ as
long as there are representatives of
all kinds of galaxy in the sample.
We have applied it to
the lowest redshift bin because the 
low lower redshift limit (0.001) means there are galaxies
with very low dust masses in the sample, which is not the case
for the higher redshift bins. However, even if the answer
is unbiased, it will be very noisy if there are only a few
representatives of a class, which is the case for galaxies
with the dust masses typical of ETGs (\S 2). This indeed is what we
see in Figure 3, where 
$N(SSFR,M_*)$ is shown as a grayscale.
The distribution, after it has been corrected for Malmquist bias, is clearly very noisy but 
lies, as expected, well below the observed distribution. 
Note that this reconstructed GS is particularly noisy at the lower right-hand
end because of the small number of ETGs detected in H-ATLAS (Section 2).

We fitted the polynomial in equation 1 to the datapoints again, this
time weighting each point by $1/V_{acc}$ in order to correct for the
effect of Malmquist bias. 
The dashed black line in Figure 3 shows the best-fitting polynomial.
As expected, it lies well below the 
polynomial that is the best fit to the unscaled datapoints. 
We have 
also plotted this Malmquist-bias-corrected line onto Figure 1, which shows the GS derived from
the HRS.
Without being a particularly good fit, the line passes through
the middle of the HRS points, showing that the GS derived from
a volume-limited survey and the GS derived from correcting a
far-infrared survey for Malmquist bias are consistent.
In an earlier paper
we showed that the GS from the HRS is consistent with the
low-redshift GS derived using other methods (Eales et al. 2017),
and the results from this section show that the GS derived from
two very different Herschel surveys are also in reasonable
agreement.

\begin{figure}
\includegraphics[trim=0mm 14mm 0mm 25mm, clip=True, width=0.49\textwidth]{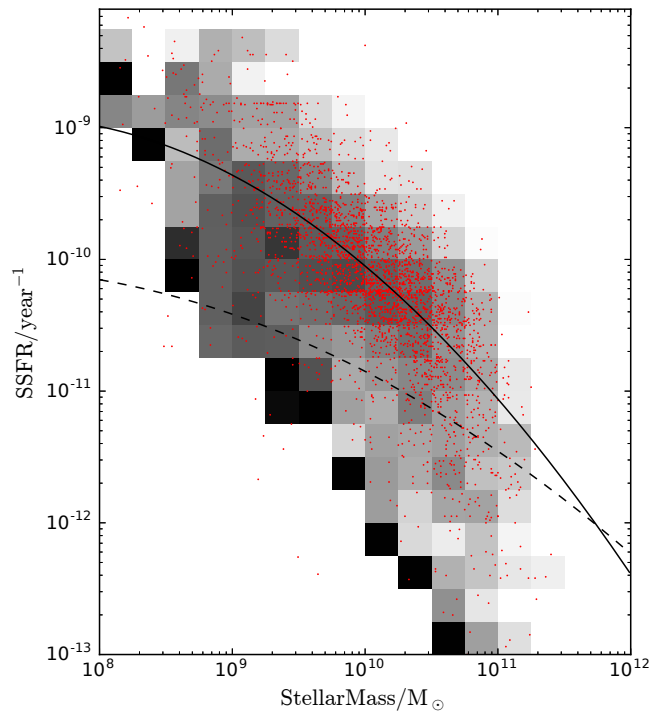}
  \caption{Specific star-formation rate versus galaxy stellar mass for the H-ATLAS
galaxies in the redshift range $0.001 < z < 0.1$. The red points
show the positions of the H-ATLAS galaxies. The grayscale shows
how the number-density of H-ATLAS galaxies
varies over the diagram after making a correction
for the
effect of accessible volume (see text).
The solid black line shows the best-fit 2nd-order polynomial to the raw data points;
the dashed line shows the fit when the data points are weighted by 1/accessible volume.
The form of the polynomial is
$\rm log_{10}(SSFR) = a + b \times (log_{10}M_* - 10.0) + c \times
(log_{10}M_*-10.0)^2$. For the raw datapoints the values of $a$, $b$ and
$c$ are -10.05, -0.85 and -0.16, respectively, and for the weighted points the values
are -10.85, -0.52 and -0.09, respectively.
}
\end{figure}

In the remainder of this section we describe 
some analysis whose original goal was to test our method for
correcting Malmquist bias but which
turned out to have an unexpected and interesting result.
We originally decided to test the method by using it to estimate
the stellar mass function for star-forming galaxies, which we
could then compare with the same stellar mass function derived
from optical surveys.

We estimated the galaxy mass function from the H-ATLAS galaxies
in each redshift bin using the following formula:

\begin{align}
\phi(M_*) dM = \sum_i { 1 \over V_{acc,i}}
\end{align}
\smallskip

\noindent In this formula, the sum is over all galaxies with $M_* < M_i < M_* + dM$,
and $V_{acc,i}$ is the accessible volume of each galaxy (equation 2).

After estimating the galaxy mass function from the H-ATLAS data, we then calculated:

\smallskip
\begin{align}
f = { \left( \phi(M_*) \right)_{submm} \over \left( \phi(M_*) \right)_{optical} }
\end{align}
\smallskip

\noindent in which the numerator is the stellar mass function derived from the H-ATLAS
galaxies using equation (4) and the denominator is the galaxy stellar mass function 
for star-forming galaxies derived from optical samples; for the latter,
at $z < 0.2$ we used the mass function from
Baldry et al. (2012) and at $z>0.2$ we used the
mass function for $0.2 < z < 0.5$ derived by Ilbert et al. (2013). 

Figure 4 shows $f$ plotted against galaxy stellar mass for the four redshift bins. 
For the two
highest redshift bins, the value of $f$ is much less than one, showing that at $z>0.2$, even after
correcting for accessible volume,
we are missing a large fraction of star-forming galaxies. 
This result is not unexpected because we showed in Section 2 that
the base sample is seriously incomplete in these bins. 

In the
second lowest redshift bin ($0.1 < z <0.2$), the H-ATLAS
mass function is incomplete at stellar masses of $<10^{10}\ M_{\odot}$,
and in the lowest redshift bin it is incomplete at stellar masses of $<10^9\ M_{\odot}$.
Above these stellar masses, however, 
$f$ reaches values that are much greater than
1, reaching values of 3-5 at the highest stellar masses.
At first sight, this result suggests that a far-infrared survey is actually
much better at finding star-forming galaxies than an optical survey, with
optical surveys missing a population of star-forming galaxies.
We will investigate this result further in the following section.

\begin{figure}
\includegraphics[trim=0mm 40mm 0mm 0mm, clip=True, width=0.49\textwidth]{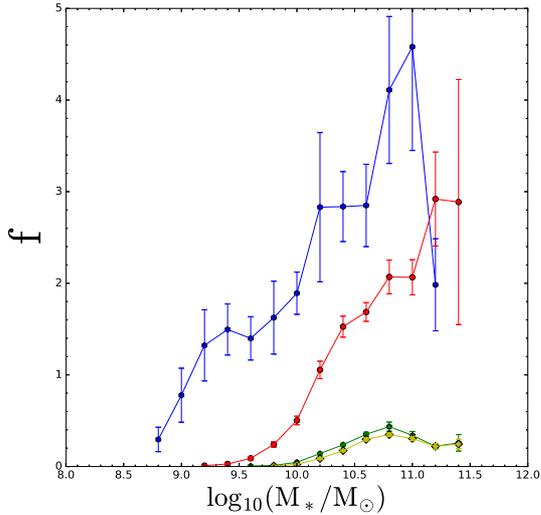}
  \caption{The ratio of the galaxy stellar mass function derived from H-ATLAS to 
that derived from optical surveys plotted against galaxy stellar mass
for four redshift bins: $0.001 < z < 0.1$ - blue symbols; $0.1 < z < 0.2$ -
red symbols; $0.2 < z < 0.3$ - green symbols; $0.3 < z < 0.4$ - light
green symbols.
}
\end{figure}

\section{Red and Blue Galaxies as seen by Herschel}

\subsection{The Galaxy Sequence}

In the comparison of the stellar mass functions at the
end of the previous section,
we implicitly assumed that the galaxies
detected by H-ATLAS are star-forming galaxies. However, there
is intriguing evidence that {\it Herschel} surveys do also detect
a population of galaxies that have red colours (Dariush et al.
2011, 2016; Rowlands et al. 2012; Agius et al. 2013). These red colours might
indicate a galaxy with an old stellar population or a star-forming
galaxy with colours reddened by dust.
In
this section and the next one, we step back from our previous
assumption
about the kind of galaxy that should be detected by a far-infrared survey;
instead we use the optical colours and spectra of the galaxies to
determine empirically what kinds of galaxy are actually detected.

\begin{figure}
\includegraphics[trim=0mm 10mm 0mm 0mm, clip=True, width=73mm]{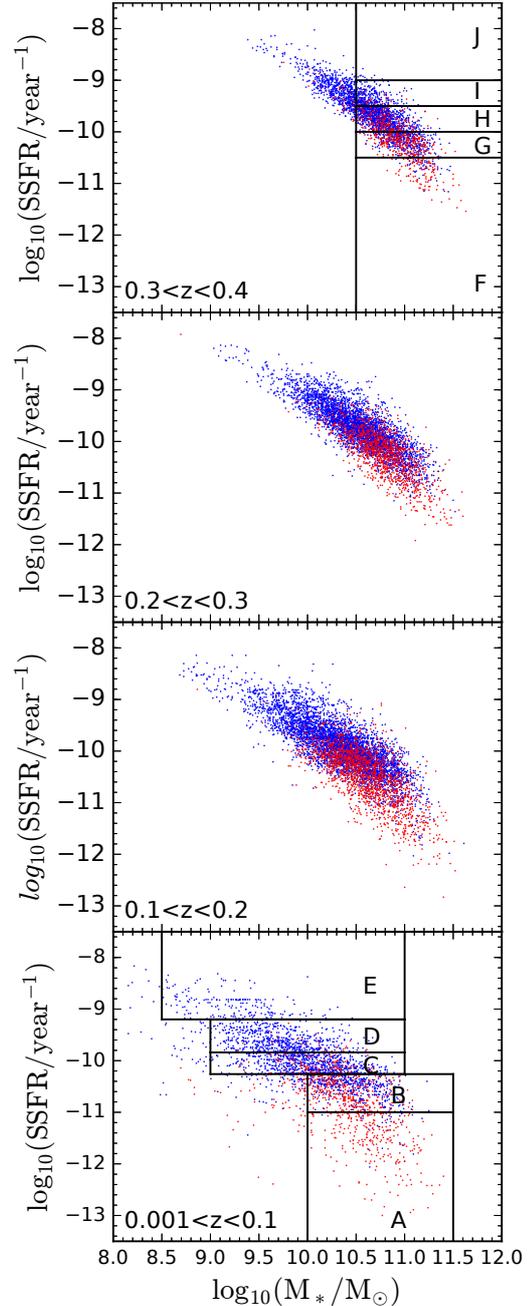}
  \caption{Specific star-formation rate versus stellar mass in the four redshift
bins.
Red points and blue points show galaxies that have redder and bluer
rest-frame $g-r$ colours, respectively, than the colour defined
by equation (6). The boxes show the ranges of SSFR
and stellar mass used to produce the stacked spectra shown in Figures 6 and
7 (\S 4.2) (use the letter in the box to find the corresponding spectrum).
}
\end{figure}

\begin{figure}
\includegraphics[trim=0mm 10mm 0mm 0mm, clip=True, width=73mm]{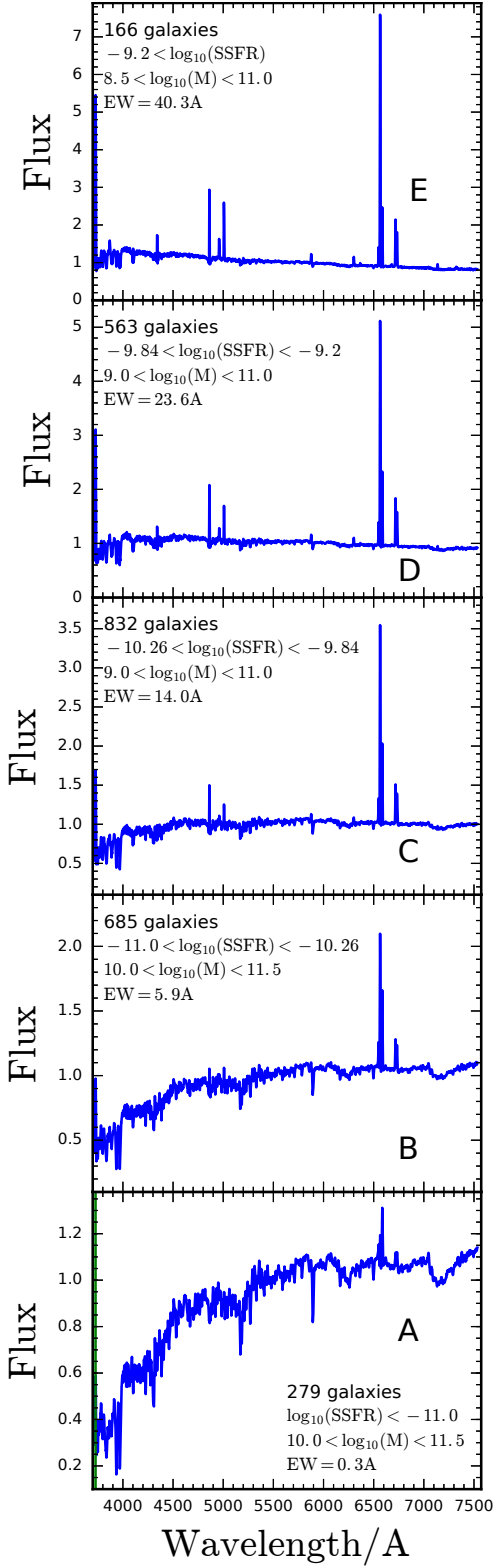}
  \caption{Median rest-frame spectra of galaxies 
in the redshift range $0.001 < z < 0.1$ in the five boxes
shown in the bottom panel of Figure 5 (use the letter to find the region).
The ranges of stellar mass and SSFR 
for each region
are
given by the spectrum.
Note (a) how, as one moves down the panels to lower values
of SSFR, the equivalent width of the H$\alpha$ line also decreases
and (b) how the red optical colours of
the galaxies in the bottom box in Figure 5 are clearly
caused by an old stellar population rather than by reddening by dust.
}
\end{figure}

\begin{figure}
\includegraphics[trim=0mm 10mm 0mm 0mm, clip=True, width=73mm]{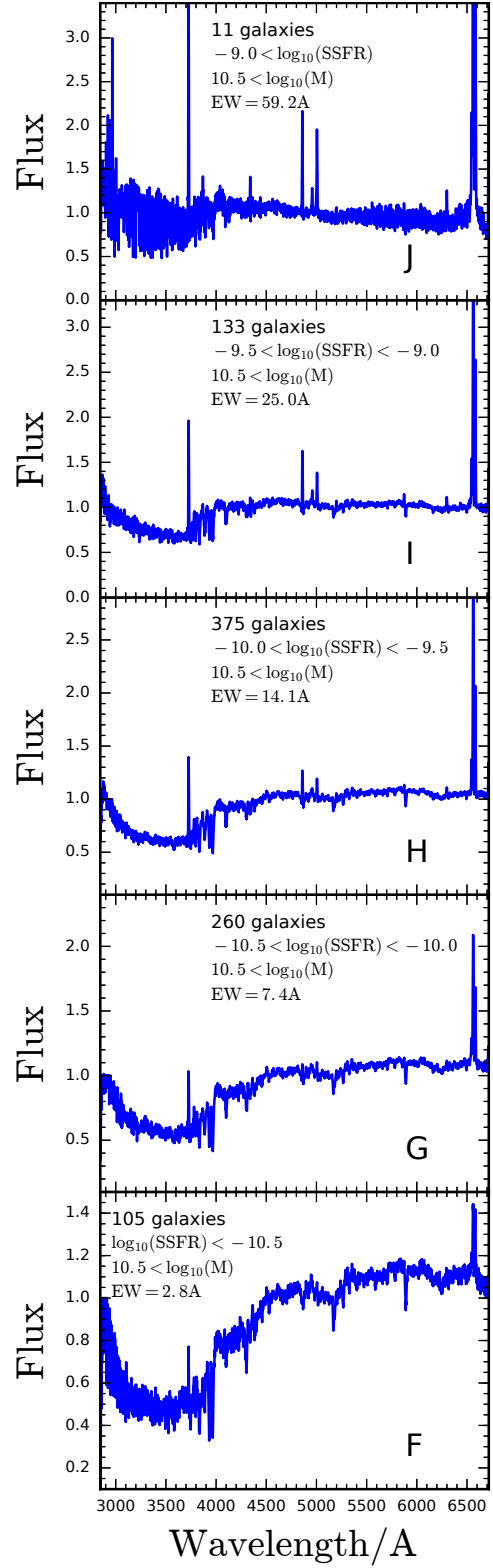}
  \caption{Median rest-frame spectra of galaxies in the redshift
range $0.3 < z < 0.4$ in the five boxes shown in
the top panel of Figure 5 (use the letter by the spectrum to find the region).
The ranges of stellar mass and SSFR 
for each region
are given by the spectrum.
Note how the spectra have both a clear 4000\AA\ break,
showing the existence of an old stellar population,
and
strong H$\alpha$ emission and a $UV$ upturn, indicating a high star-formation
rate.
}
\end{figure}

In our investigation we have used the optical colours to
separate the H-ATLAS galaxies into two classes 
using two alternative criteria from Baldry et al.
(2012). Baldry et al. called these classes `star-forming' and `passive',
but we will call them `blue' and `red', since the former nomenclature
makes the assumption that red galaxies are not forming stars.
As the first criterion we use the rest-frame $g-r$ colour of the
galaxy to classify it as red or blue using the dividing
line on the colour versus absolute magnitude diagram:

\smallskip
\begin{align}
g-r = -0.0311 M_r + 0.0344
\end{align}
\smallskip

\noindent As the second criterion we use the rest-frame $u-r$ colour
with the dividing line on the colour versus absolute magnitude diagram
being:

\smallskip
\begin{align}
u-r = 2.06 - 0.244 {\rm tanh}\left( {M_r + 20.07 \over 1.09} \right)
\end{align}
\smallskip

\noindent We calculated the rest-frame colours by calculating individual
k-corrections for each galaxy
by applying $KCORRECT\ v4\_2$
(Blanton et al. 2003; Blanton and Roweis 2007).
In brief, this package finds the linear combination of
five template spectra
that gives the best fit to the five SDSS magnitudes for each
galaxy and then uses this model to calculate the K-correction
for the galaxy
(Blanton and Roweis 2007).
Some additional details of the implementation of the
code are given in Loveday et al.
(2012).

Equations 6 and 7 were determined by Baldry et al. (2012) from
the low-redshift galaxy population. However, even a galaxy 
today
in which no stars have formed for the last 10 billion years
will have had bluer colours in the past because of the evolution
in the turnoff mass on the stellar main sequence.
We therefore
investigated the effect of adding a small correction to
these equations to model the expected evolution in the colours
of a very old stellar population. 
Our model of this effect was
based on a model of a single stellar population
from Bruzual and Charlot (2003) with a Salpeter
initial mass function and solar metallicty. We assumed
that the galaxy started forming 12 Gyr ago with
the star-formation rate
proportional to $exp(-t/\tau)$ and $\tau = 1\ Gyr$.

Table 2 gives the percentages of red galaxies in the different
redshift bins for the two colour criteria and also shows the effect of
adding the evolutionary correction. 
Rather surprisingly, even without making the 
evolutionary correction, $\simeq$15-30\% of the H-ATLAS galaxies are red galaxies.
This
is higher than the value of $\simeq4.2\%$ found by Dariush et al. (2016)
for H-ATLAS galaxies at $z<0.2$. We suspect that the difference arises
because
Dariush et al. used optical-UV colours 
rather than optical colours, a suspicion we will explain in the next
section (\S 4.2).
Figure 5 shows the GS again, but this time with the points colour-coded
to show which galaxies are red and blue according to equation 6 (equation
7 produces a very similar figure).

The significant fraction of red galaxies 
explains the values of $f>$1 in Fig. 4,
because these galaxies would have been classified
as passive galaxies
using optical criteria and so
would have been omitted from the mass function
for star-forming galaxies derived from optical surveys.
However, these red galaxies
still have significant reservoirs
of interstellar gas (after all, they are detected because
of the continuum emission from interstellar dust)
and the MAGPHYS results imply they are still forming stars.
We will investigate further the properties of this
interesting population in the next section.
Figures 5 is a striking demonstration of why the GMS produced from a subset
of galaxies classified as star-forming will generally be flatter
than the GS we have derived from the two {\it Herschel} surveys. 
Imagine removing all the red galaxies
in Figures 5;
the GMS would then have
a much flatter slope.

Although there are fewer optically-red galaxies than optically-blue
galaxies in H-ATLAS, a simple argument shows that optically-red star-forming
galaxies are not a peripheral population.
The optically-red galaxies are under-represented in H-ATLAS
because of Malmquist bias.
At a given stellar mass, Figure 5 shows
that optically-red galaxies generally have lower values of SSFR than optically-blue
galaxies, which in turn means a lower star-formation rate and, through the
Kennicutt-Schmidt law, a lower gas and dust mass - leading to
a smaller accessible volume and Malmquist bias.
Figure 4, where we have attempted
to correct for Malmquist bias, shows this quantitatively.
The implication of the figure is that, at a given stellar mass, the space-density
of optically-red star-forming galaxies is at least as high
as that of optically-blue star-forming galaxies.

Optical investigators have generally missed this population
because they classify these galaxies as passive.
However, they have the following excuse.
A comparison of the stellar mass
functions given by Baldry et al. (2012; their
Figure 15) for star-forming galaxies and passive galaxies (precisely equivalent to
our optically-blue and optically-red classes) shows that
the space density of the two galaxy types
is the same at a stellar mass of $\simeq10^{10}\ M_{\odot}$, but that
at higher stellar masses the space-density of passive galaxies is higher, with
a maximum difference of a factor of $\simeq5$
at a stellar mass of $\simeq 4 \times 10^{10}\ M_{\odot}$.
Given the much larger number of passive galaxies, even
a small change in
how one divides
galaxies into the passive and star-forming classes
will have a large effect on estimates of the space-density of star-forming
galaxies.

Not all optical investigators, however, have missed this population. In their
elegant reanalysis of the SDSS galaxy sample, O17 showed there is an intermediate
population of galaxies between those that are rapidly forming stars and passive
galaxies. The galaxies in this intermediate class
are still forming stars and seem identical to our
optically-red star-forming galaxies.
O17 conclude that at a given stellar
mass the number of galaxies in this intermediate class is roughly the same
as the number in the rapidly star-forming class, thus reaching exactly the
same conclusion as we do but starting from a traditional optical survey. 

\begin{table}
\caption{Percentages of red galaxies in H-ATLAS}
\begin{tabular}{ccccc}
\hline
Redshift & (g-r) & (u-r) & (g-r) & (u-r) \\
&  & & plus evolution & plus evolution \\
\hline
$0.001 < z < 0.1$ & 27\% & 16\% & 29\% & 18\% \\
$0.1 < z < 0.2$ & 26\% & 15\% & 31\% & 18\% \\
$0.2 < z < 0.3$ & 24\% & 18\% & 33\% & 23\% \\
$0.3 < z < 0.4$ & 21\% & 21\% & 35\% & 28\% \\
\end{tabular}

\medskip

The percentage of H-ATLAS galaxies
in different redshift ranges classified as red using equation
6 (columns 2 and 4) and equation 7 (columns 3 and 5). In columns
4 and 5 we add a correction to equations 6 and 7 to allow
for the expected evolution of a very old stellar population (see text).
\end{table}

\subsection{Stacking spectra - the nature of the red and blue galaxies}

The colours of the optically-red H-ATLAS galaxies might indicate an old stellar
population or alternatively a star-forming galaxy whose colours are reddened by dust.
We attempted to distinguish between these possibilities using the galaxies' spectra.
The spectra come from the GAMA and SDSS projects, with most of the spectra
coming from the former. Hopkins et al. (2013) describe the calibration and
other technical details of the GAMA spectra. The GAMA project used
the AAOmega spectrograph on the AAT, which has fibres with
an angular diameter on the sky of 2 arcsec.
This corresponds to a physical size at
redshifts
of 0.1, 0.2, 0.3 and 0.4 of 3.6 kpc, 6.6 kpc, 8.9 kpc and 10.7 kpc, respectively.

We divided
the SSFR versus stellar mass diagrams for the
H-ATLAS galaxies with $0.001 < z < 0.1$ 
and with $0.3<z<0.4$ each into five boxes, which are shown in the
bottom and top panels of Figure 5. 
We then calculated
the median rest-frame spectrum of all the GAMA and SDSS
spectra in each box (Figures 6 and 7).
We measured
the equivalent width of the H$\alpha$ line from each spectrum, using
the wavelength ranges $6555.5 < \lambda < 6574.9\AA$ to measure the flux
in the line and the wavelength ranges
$6602.5 < \lambda < 6622.5\AA$ and
$6509.5 < \lambda < 6529.5\AA$ to estimate the mean value of the continuum
at the line wavelength. The H$\alpha$ equivalent width,
the ranges of stellar mass and SSFR for each box, and the number of galaxies
in the box are shown by the side of each spectrum in Figures 6 and 7.

First, let us consider the stacked spectra for the low-redshift
galaxies (Figure 6).
The galaxies in the bottom box in Figure 5 almost all have red optical
colours.
The stacked spectrum for this box, which is shown in the lowest panel
in Figure 6,
is strongly characteristic of an old stellar population. The red colours 
are therefore generally the result of the age of the 
stellar population rather than dust reddening.

This figure gives further insights into why different studies of the GMS can find very
different results. 
As we move down the panels in Figure 6, the appearance of the
stacked spectra gradually changes, with the
equivalent width of the H$\alpha$ line, the brightest
emission line in the spectra, steadily decreasing.
This is not surprising (although it is reassuring) because the luminosity
of the H$\alpha$ line is often used to estimate a galaxy's  star-formation rate
(Davies et al. 2016; Wang et al. 2016b).
When the H$\alpha$ line is used in studies of the GMS to separate
star-forming from passive galaxies, the dividing line is
usually an H$\alpha$ equivalent width in the range $3\AA < EW < 10\AA$ (Bauer et al.
2013; Casado et al. 2015). The equivalent
width
of H$\alpha$ 
in the three lowest boxes in Figure 6 is, 
in order of increasing SSFR, 0.3, 5.9 and 14.0\AA. Therefore,
when this method is used to separate
star-forming and passive galaxies, the shape of the GMS that is found depends critically
on the exact value of the equivalent width used to divide the galaxies.

These results also suggest something more fundamental 
which we will return to later. Whereas the optical view
of the galaxy population is that there are two distinct classes
of galaxy (\S 1), the {\it
Herschel} results show more continuity.
The red galaxies have colours and stacked spectra
that imply they have old stellar populations but their detection by {\it
Herschel} shows
they contain a substantial ISM
 - and both our MAGPHYS results and the
H$\alpha$ equivalent widths imply they are still forming stars.
Furthermore, the overlap of red and blue galaxies in Figure 5 also
implies that the two classes are not clearly physically distinct.

The blurring between the two classes is even more evident when we turn to the
high-redshift population. In Figure 7 we 
show the result of stacking the spectra of the
galaxies in the redshift range $\rm 0.3 < z < 0.4$ and $\rm log_{10}(M/M_{\odot}) > 10.5$.
The base sample is highly incomplete in this redshift range, although
the incompleteness is most severe at lower stellar masses (Section 2). 
The stacked spectra for this high-redshift bin are visually quite startling
because they all show clear evidence of both an
old and a young stellar population. In all the stacked
spectra, there is a clear 4000\ \AA\ break,
evidence of an old stellar population. Since the SDSS u-band is centred at
$\simeq$3500\ \AA, the existence of the 4000 \AA\ break immediately explains why
so many of the galaxies fall into the optically-red class. However, in all the stacked
spectra there are also clear
signs of a high star-formation rate, including strong emission lines, in particular
H$\alpha$, and a $UV$ upturn.
We don't know whether this $UV$ upturn is present in the galaxies in the low-redshift bin
because our spectra for this bin (Figure 6) do not extend to a low enough rest-frame wavelength,
but a $UV$ upturn would explain why Dariush et al. (2016) found a much smaller
fraction of red galaxies when using the $UV$-optical colours to classify
the galaxies.

Figure 7 shows that the rapid low-redshift evolution that we have seen
in previous H-ATLAS studies (\S 1) is caused by galaxies with high stellar masses and
a large population of old stars. In the Universe today, galaxies like this are forming
stars at a very low rate but four billion years back in time 
they were clearly forming stars at a much faster rate.

\subsection{The evolution of the red and blue galaxies}

In earlier papers (Dye et al. 2010; 
Wang et al. 2016a), we showed that the
H-ATLAS 250-$\mu$m luminosity function shows rapid evolution
over the redshift range $0 < z < 0.4$ with significant evolution even
by a redshift of 0.1. Marchetti et al.
(2016) found similar results from an analysis of the results of
the other large {\it Herschel} extragalactic survey, HerMES.
In this section we
consider the evolution of the 250-$\mu$m luminosity function separately for red
and blue galaxies.

In this case we started with {\it all} the galaxies
from the GAMA fields
detected at $>4\sigma$ at 250 $\mu$m with spectroscopic (by preference)
or photometric redshifts in the range $0.001 < z < 0.4$
(Valiante et al. 2016; Bourne et al. 2016). 
Of the 25,973 H-ATLAS galaxie in this sample,
20,012 have spectroscopic
redshifts.
We used the $u-r$ colour criterion (equation 7) to divide the H-ATLAS
galaxies into red and blue galaxies, although the results using the $g-r$ colour
criterion (equation 6) were very similar. 
We made the small correction to equation 7 that allows for the fact that the
colours of even a very old stellar population must have been redder in the
past (Section 4.1), although this actually makes very little difference to the results.

To calculate the luminosity function for each class, we used the estimator invented by
Page and Carerra (2000), since this has some advantages for
submillimetre surveys (Eales et al. 2009):

\smallskip
\begin{align}
\phi(L_1<L<L_2,z_1<z<z_2)\Delta log_{10}L\Delta z = n/V
\end{align}
\smallskip

\noindent in which $n$ is the number of galaxies in this bin of luminosity
and redshift and $V$ is the accessible volume averaged over
the luminosity range of this bin.
We multiplied the luminosity function in each
redshift bin by  
$ 1/C$, where $C$ is the 
estimated efficiency of our method for
finding the galaxies producing the
far-infrared emission (Bourne et al. 2016; column 3 of Table 1).

Figure 8 shows the luminosity function for the optically-red and optically-blue galaxies.
Strong evolution is seen in the luminosity function
for both populations, with the evolution in the red population looking
slightly stronger.
We quantified the evolution by fitting a Schechter function to
each empirical luminosity function. 
For the lowest-redshift luminosity function, 
we allowed all three parameters
of the Schechter function - $\alpha$, $L_*$ and $\phi_*$ - to
vary.
For the other luminosity functions, which have a smaller range of
luminosity, we only allowed $L_*$ and $\phi_*$ to vary,
using the value of $\alpha$ from the low-redshift bin.
We then used the estimates of the parameters at each redshift
to investigate the evolution of
$\phi_*$ and $L_*$.
We assumed that the
evolution has the form $\phi_* = \phi_{*0} (1+z)^n$
and $L_* = L_{*0} (1+z)^m$. We found
$n=0.24\pm0.04$ and $m=3.69\pm0.01$ for the optically-blue
galaxies and $n=3.66\pm0.06$ and $m=1.86\pm0.01$ for the
optically-red galaxies.
Therefore, there is strong evolution
in $L_*$ and $\phi_*$ for the red galaxies
and strong evolution in $L_*$ for the blue galaxies.

Bourne et al. (2012) also found evidence
for evolution in submillimetre luminosity and dust mass
for both red and blue galaxies, with marginal evidence
of stronger evolution for the red galaxies. 
The stacking analysis of Bourne et al.
was carried out on an optically-selected sample of galaxies,
and is therefore evidence that the 
strong evolution we see for optically-red galaxies
is not just a phenomenon associated with an interesting, but ultimately
unimportant population detected by
{\it Herschel} but applies to the whole galaxy population.

\begin{figure}
\includegraphics[width=74mm]{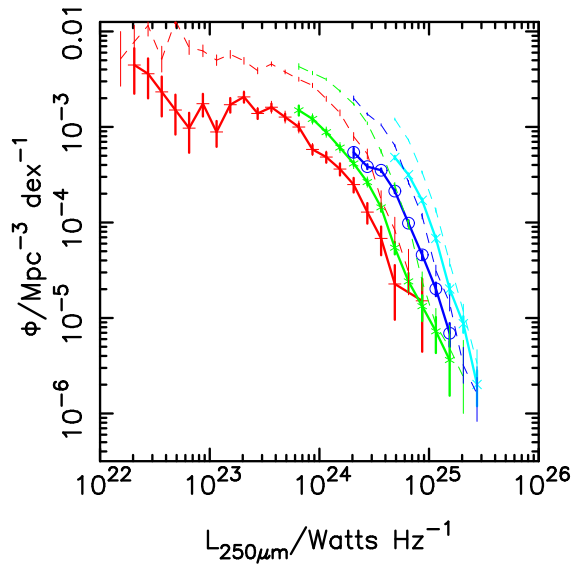}
  \caption{The 250-$\mu$m luminosity functions for the optically-red
H-ATLAS
galaxies (solid lines) and
optically-blue H-ATLAS galaxies (dashed lines).
The colours correspond to the following redshift
ranges: red - $0<z<0.1$; green - $0.1<z<0.2$;
dark blue - $0.2<z<0.3$; light blue - $0.3<z<0.4$.
}
\end{figure}

\subsection{How star-formation efficiency varies along the GS}

Results from the {\it Herschel} Reference Survey show that galaxy
morphology changes gradually along the GS, the morphologies
moving 
to earlier types
on the Hubble sequence as one moves down the
GS (Fig. 1; Eales et al. 2017).
This progression implies an increase in the bulge-to-disk ratio,
which Martig et al. (2009) have argued should lead to a decrease
in star-formation efficiency (SFE, star-formation rate divided by
ISM mass). 
In this section we test this hypothesis by investigating
whether SFE varies along the GS.
There is already some evidence from other surveys that SFE
and SSFR are correlated
(Saintonge et al. 2012; Genzel
et al. 2015).

We have restricted our analysis to the H-ATLAS galaxies in the
redshift range $0.001 < z < 0.1$.
In our analysis we use the MAGPHYS estimates of the star-formation
rate and the dust mass, using the dust mass of each galaxy to estimate
the mass of the ISM. Many authors
(Eales et al. 2012; Scoville et
al. 2014; Groves et al. 2015; Genzel et al. 2015) have 
argued 
this is a better way of estimating the ISM mass than the
standard method of using the 21-cm and CO lines, because of the many
problems with CO, in particular
the evidence that one third of the molecular gas in the Galaxy contains no
CO (Abdo et al. 2010; Planck Collaboration 2011; Pineda et al.
2013), which is probably because of photodisintegration of the
CO molecule.
Dust grains, on the other hand, are quite robust,
and the main problem with the dust method is the fact that the dust-to-gas
ratio is likely to depend on metallicity, a problem of course that
is shared by the CO method.

There is a lot of evidence that above a transition 
metallicity ($\rm 12+log(O/H) \simeq 8.0$) 
the dust-to-gas ratio is proportional to the metallicity (James et al. 2002;
Draine et al. 2007; Bendo et al. 2010; Smith et al. 2012c; Sandstrom
et al. 2013; R\'emy-Ruyer et al. 2014).
In order to test how robust our results are to
the metallicity correction, we have used three
different methods for doing this correction.
In the first method
we make no correction for metallicity and assume
that each galaxy has a dust-to-gas ratio of 0.01.
In the second method
we estimate the metallity of each galaxy from its stellar
mass using the relationship 
found by Tremonti et al. (2004):

\smallskip
\begin{align}
12 + log_{10}(O/H) = -1.492 & + 1.847(log_{10}M_*) \nonumber\\
& - 0.08026(log_{10}M_*)^2
\end{align}
\smallskip

\noindent 
We then assume that the dust-to-gas ratio is proportional to the
metallicity and that a galaxy with solar metallicity
has a dust-to-gas ratio of 0.01.
The third method is the same as the
second
except that we use the relationship
found by Hughes et al. (2013) from their study of HRS galaxies:
\smallskip
\begin{align}
12 + & log_{10}(O/H) = 22.8 - 4.821(log_{10}M_*) \nonumber\\
& +0.519(log_{10}M_*)^2 - 0.018(log_{10}M_*)^3
\end{align}
\smallskip

The obvious statistical test is to see
whether there is any correlation between SFE
and SSFR. However, this
is dangerous because both quantities
are ratios with
the star-formation rate in the numerator; thus errors
in the star-formation-rate estimates may lead to
a spurious
correlation.
Instead, we have compared the SFE
of
the optically-red and optically-blue galaxies, since this classification
was done without using any of the MAGPHYS estimates and yet we
know from Figure 5 that optically-red galaxies have lower SSFR values
than optically-blue galaxies.

Figure 9 shows the distributions of SFE
for the optically-red and optically-blue galaxies when
the metallicity correction is made using equation 9.
The
figures produced using the two other methods look very
similar. 
The mean values of SFE for
the optically-red and optically-blue galaxies are given in Table
3, for all three
metallicity-correction methods
and for both methods of
classifying the galaxies as red or blue.
In each case, we have
compared the two distributions using the two-sample
Kolmogorov-Smirnov test, testing the null hypothesis
that the two samples are drawn from the
same population.
The values of the KS statistic and the probabilities that the two samples
are drawn from the same population 
are given 
in Table 3.

\begin{table*}
\caption{The star-formation efficiency of optically-red and optically-blue
galaxies}
\begin{tabular}{cccccccc}
\hline
Colour & metallicity & $n_{red}$ & $n_{blue}$ & $<SFE_{red}>$ & $<SFE_{blue}>$ & KS & Prob. \\
 & correction & & & & & & \\
\hline
$g-r$ & None & 885 & 2502 & -9.59$\pm$0.01 & -9.19$\pm$0.01 & 0.40 & $<<$0.1\%\\
$g-r$ & Tremonti  & 885 & 2502 & -9.24$\pm$0.01 & -8.95$\pm$0.01 & 0.31 & $<<$0.1\% \\
$g-r$ & Hughes  & 885 & 2502 & -9.74$\pm$0.01 & -9.41$\pm$0.01 & 0.35 & $<<$0.1\% \\
$u-r$ & None & 534 & 2853 & -9.68$\pm$0.02 & -9.22$\pm$0.01 & 0.43 & $<<$0.1\% \\
$u-r$ & Tremonti  & 534 & 2853 & -9.33$\pm$0.02 & -8.97$\pm$0.01 & 0.38 & $<<$0.1\% \\
$u-r$ & Hughes  & 534 & 2853 & -9.84$\pm$0.02 & -9.43$\pm$0.01 & 0.42 & $<<$0.1\% \\
\hline
\end{tabular}

\medskip

The columns are as follows: Col. 1--the colour used to
divide galaxies into optically-red and optically-blue galaxies; col. 2--the method used
to correct the dust-to-gas ratio for the effect
of metallicity (see text for details); col. 3--the number of optically-red
 galaxies;
col. 4--the number of optically-blue galaxies; col. 5--the mean value of the logarithm of
star-formation
efficiency for the optically-red galaxies;
col. 6--the mean value of the logarithm of star-formation efficiency for the
optically-blue
galaxies; col. 7--the value of the two-sample Kolmogorov-Smirnov statistic used
to compare the SFE distributions of the optically-red and optically-blue
galaxies;
col. 8--the probability that such a high value of the KS statistic would be
obtained if the two distributions were drawn from the same population.
\end{table*}

The KS test shows that for all six variations on the basic
method the
probability that the star-formation efficiency of the
optically-red and optically-blue galaxies is the same is $<<0.1\%$.
Table 3 shows that the
mean SFE
of the optically-red galaxies is lower by a factor of $\simeq1.9-2.9$ than
the optically-blue galaxies. Saintonge et
al. (2012) found that as the SSFR decreases by 
a factor of $\simeq$50, SFE
decreases by a factor of $\simeq$4 (see their Fig. 2).
The shift in SSFR between the optically-blue and
optically-red galaxies in Figure 5 is a little less than this
and so our result seems 
to be in reasonable agreement with
their result.

\begin{figure}
\includegraphics[width=74mm]{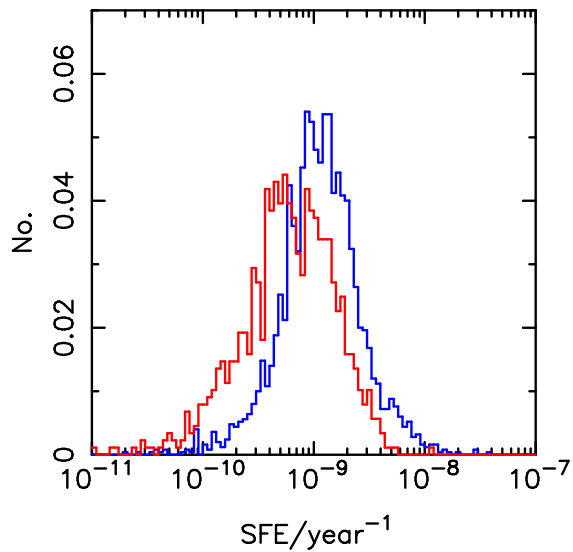}
  \caption{{Histograms of star-formation efficiency for the
optically-red galaxies (red line) and optically-blue galaxies
(blue line) for the H-ATLAS galaxies in the
redshift range $0.001 < z < 0.1$.
In this diagram, we have used equation 6 ($g-r$) to split the
galaxies into the two classes 
and the metallicity relationship of Tremonti et al. (2004)
to correct for the metallicity effect (see text).
The histograms have been normalised
so that the areas under both are the same.}
} 
\end{figure}

\section{Discussion}

\subsection{What have we learned from Herschel?}

We have learned three main things from the {\it Herschel} extragalactic surveys
that need to be
explained by any comprehensive theory of galaxy evolution.

We have learned first that the galaxy population shows rapid
evolution at a surprisingly low redshift. This is evident in the submillimetre
luminosity function (Dye et al. 2010; Wang et al. 2016a), the dust masses of galaxies
(Dunne et al. 2011; Bourne et al. 2012) and the star-formation rates, whether estimated
from radio continuum observations (Hardcastle et al. 2016) or from
bolometric dust emssion (Marchetti et al. 2016).
The rapid evolution in dust mass and star-formation rate are of course
connected,
since stars form out of gas and the dust traces the interstellar
gas reservoir. 

Several recent studies have found
that the GS is curved, whether only star-forming
galaxies are plotted (Whitaker et al. 2014; Lee et al. 2015; Schreiber
et al. 2016; Tomczak et al. 2016) or all galaxies are plotted
(Gavazzi et al. 2015; O17).
The results from the two {\it Herschel} surveys confirm and
extend this result.
Both the {\it Herschel} surveys, selected in very different
ways, show that galaxies lie on an extended curved GS 
rather than a star-forming GMS and a separate region of `passive' or `red-and-dead'
galaxies (Figures 1 and 3).
The GS shown in Figures 1 and 3 extends down to
lower values of SSFR than 
most of the other studies
because we have made no attempt to remove `passive' galaxies, a
distinction which we argue below is anyway rather meaningless.

Third, we have learned that the operational division often used in optical studies
between red and blue galaxies looks distinctly arbitrary when viewed from a submillimetre
perspective. 15-30\% of the H-ATLAS galaxies fall into the red category, but these
galaxies clearly are not quenched or passive galaxies, since they still have large
reservoirs of interstellar material and are still forming stars.
The distinction looks even less clearcut at $z\simeq0.3-0.4$, where
the spectra of both red and blue galaxies are qualitatively very similar (Fig. 7).
After correcting for the effect of Malmquist bias in H-ATLAS, we 
find that, at a constant stellar
mass the space-density of these optically-red star-forming galaxies, which are
missing from most optical studies, is at least as high as the space-density
of the optically-blue star-forming galaxies that are included in optical studies.
O17 reached a similar conclusion from their reanalysis of the optically-selected
SDSS galaxy sample.

The combination of the second and third results point to a very
different picture of the galaxy population than the standard
division into 
two dichotomous classes - 
whether one calls these classes `late-type' and `early-type', `star-forming' and 
`passive', `star-forming'
and `red-and-dead', or `star-forming' and `quenched'. 
The smooth appearance of the GS when all galaxies are plotted rather
than just galaxies 
classified as `star-forming' (Figs 1 and 3), the gradual change in galaxy
morphology along the GS (Fig. 1), the fact that the average galaxy spectra
change gradually along the GS (Figs 6 and 7), the strong evolution shown by
both optically-red and optically-blue galaxies - all of these suggest that galaxies are better
treated as a unitary population rather than two dichotomous classes.  
As we discuss in \S 5.5, there is actually a a lot
of recent evidence from observations
in other wavebands that leads to the same conclusion.

If our
conclusion is correct, that galaxies form a unitary population rather
than two dichotomous classes, much of the rationale
for rapid quenching models (\S 1) vanishes.

Finally, in this section, we consider the identity of the
galaxies producing the rapid low-redshift evolution.
The smoking gun is possibly Figure 7.
The stacked spectra in this figure show that
the high-redshift H-ATLAS galaxies both contain an old stellar
population, shown by the significant 4000\ \AA\ spectral
break (and the red optical colours), but are also forming stars at a high rate,
shown by the ultraviolet emission and strong emission lines.
The galaxies used to produce these average spectra all have stellar masses
$\rm >10^{10.5}\ M_{\odot}$. 
Today most galaxies with stellar masses
above this limit are ETGs (Fig. 1).
Therefore it seems likely that the H-ATLAS galaxies are
the fairly recent ancestors of some part of the ETG population in
the Universe today.

Arguments based on chemical abundances imply that at least 50\% of the
stellar mass of ETGs formed over $\simeq$8 billion years
in the past (Thomas et al. 2005),
but our
results and those of Bourne et al. (2012) suggest that even looking back a few
billion years is enough to see significantly enhanced star-formation rates
in the ETG population.

\subsection{Comparision with theoretical models}

The recent large-area galaxy surveys, H-ATLAS in the far-infrared and GAMA in the
optical, have made it possible to observe galaxy evolution, with good time
resolution, over the last few billion years.
The models have not kept pace with the observations and
most theoretical papers make predictions over a longer time period
with much coarser time resolution. It is therefore not possible to make a 
definitive comparision of our new results with the results from 
numerical 
galaxy simulations such as Illustris and EAGLE, although we do 
make
a comparision with the predictions in a very recent EAGLE paper at
the end of this section. We have therefore  mostly used an
analytic galaxy-evolution model to try to reproduce our results, which does have
the advantage that it is generally easier to diagnose
why an analytic model
does not match
the results than is the case for a numerical model.

Given the clear evidence that the increased star-formation rate in galaxies at
high redshift is because they contain more
gas, both evidence from our results (Dunne et al. 2011) and from those
of others (Genzel et al. 2015; Scoville et al. 2016), it
is reasonable to assume that,
to first order, galaxy evolution is governed by the supply of gas.
A useful model based on this assumption, which we will use
in this section, is the `bathtub model' or `gas regulator model'
(Bouch\'e et al. 2010; Lilly et al. 2013; Peng and Maiolino 2014).
We will use the analytic version of the bathtub model described
by Peng and Maiolino (2014; henceforth PM).

The PM model is based on three assumptions:
(a) the star formation rate is proportional to the mass of 
gas in the galaxy; (b) 
the rate at which gas is flowing out of the galaxy is proportional to the star-formation
rate; (c) the rate at which gas is flowing into the galaxy is proportional
to the growth rate of the dark-matter halo in which the galaxy is located.
The final assumption can be represented by the following equation:

\smallskip
\begin{align}
\Phi  \propto {dM_{halo} \over dt}
\end{align}
\smallskip

\noindent in which 
$\Phi$ is the gas inflow rate.
In their model, PM used the specific
mass-increase rate for a halo found in the hydrodynamic simulations
of Faucher-Giguere et al. (2011):

\smallskip
\begin{align}
<{1 \over M_{halo}} {dM_{halo} \over dt}> = & 0.0336(1+0.91z) \left( {M_{halo} \over 10^{12} M_{\odot}} \right)^{0.06}\nonumber\\
 & \times \sqrt{ \Omega_M (1+z)^3 + \Omega_{\Lambda}}\ Gyr^{-1}
\end{align}

We will examine whether this model can explain two of the
key {\it Herschel} results. First, let us consider the shape of the GS.
In their model, PM show that the equilbrium value of the SSFR of a galaxy is
given by:

\smallskip
\begin{align}
SSFR = {1 \over t - t_{eq}}
\end{align}
\smallskip

\noindent in which 
$t_{eq}$ depends on parameters such as the mass-loading factor for
outflows, the fraction of the mass of newly formed stars
that is eventually returned to the ISM, and the star-formation
efficiency. With the possible exception of the last
(\S 4.4), there is no obvious reason why any of these should
depend on stellar mass, and so the model predicts that the SSFR should
also be independent of stellar mass.
Although this
is consistent with the relatively flat GMS found in some optical studies (e.g. Peng et
al. 2010), it is clearly completely inconsistent with our results 
and the other recent findings that the GMS is strongly curved
(Whitaker et al. 2014; Lee et al. 2015; Gavazzi
et al. 2015; Schreiber et al. 2016; Tomczak et al. 2016; O17).

Now let us try to reproduce the rapid low-redshift evolution. 
We will try to reproduce both the evolution in the position
of the GS seen in Figure 2 and the evolution in the mass
of the ISM in galaxies found by Dunne et al. (2011).

Although the two highest redshift bins in Figure 2 are highly incomplete,
our 
completeness analysis (\S 3.2) showed
that at $z < 0.2$ H-ATLAS is fairly complete 
for stellar masses $\rm >10^{10}\ M_{\odot}$ once
a correction has been made for Malmquist bias (Figure 4).
We therefore fitted a second-order polynomial to
the datapoints in the two lowest redshift bins in Figure 2 for galaxies
with masses $\rm >10^{10}\ M_{\odot}$.
We found that
the mean value of SSFR 
increases by
a factor of $\simeq$2.7 between the two bins if no correction is
made for Malmquist bias and $\simeq$3.0 if a correction is made.
We can use equation 13 to predict how rapidly the mean SSFR should change
between bins according to the PM model. If we assume that the galaxy
formed 12 Gyr ago and that $t_{eq}$ is insignificant, the predicted change in SSFR between a redshift
of 0.05 and 0.15 is only 1.12, much less than the observed
evolution. 

Now let us consider the change in
the mass of the ISM.
Dunne et al. (2011) found that by a redshift of $\simeq0.45$ galaxies
contain roughly 5 times as much dust (and therefore gas\footnote{We assume
that the gas-to-dust ratio does not evolve significantly.}) as
in the Universe today. 
In the PM model, the equibrium gas mass is 
proportional to the gas inflow rate. 
On this assumption and using equations 11 and 12,
we calculate that
the gas mass should increase by a factor of $\simeq1.3$ from
a redshift of 0.05 to 0.45.

In both cases, the PM model predicts much weaker evolution than
we observe. The physical reason for this is the assumption that
the rate of increase of gas flow into a galaxy is proportional to the
rate of increase of the mass of the surrounding dark-matter halo (equation 11),
since the growth in the masses of dark-matter halos is very slow at low
redshift (equation 12). This will be a 
problem for any model, analytic or otherwise, in which the gas flow into
a galaxy is proportional to the growth rate of the surrounding halo.

Although it is not yet possible to make a definitive comparision
with the predictions of 
of the numerical galaxy simulations, we will make
a first attempt to see whether these might reproduce the
rapid low-redshift evolution 
using the recent predictions for the
star-formation rate function (SFRF; the space-density
of galaxies as a function of star-formation rate) 
by the EAGLE team
(Katsianis et al.
2017). Katsianis et al. show predictions for the SFRF at two redshifts
in the redshift range of interest: $z = 0.1$ and $z = 0.4$.
Inspection of their Figure 2 shows that the SFRF has a similar shape
at the two redshifts but is shifted upwards in star-formation rate
by a factor of $\simeq$1.55 from the lower to the higher redshift.
This is quite similar to the way the submillimetre
luminosity function evolves, because while its
shape stays roughly the same it
moves
gradually to higher luminosity as the redshift
increases
(Figure 8 of this paper; Dye et al. 2010;
Wang et al. 2016a; Marchetti et al. 2016). We can make a rough comparison
of the observations and the model if 
we make the assumption that the
characteristic luminosity of the best-fit Schechter
function ($L_*$) is proportional to
the star-formation rate. Wang et al.
(2016a) and Marchetti et al. (2016)
find, respectively, that $L*$ at 250 $\mu$m increases by a factor
of 3.25 and 3.59 over this redshift range.
This is much faster than the evolution predicted by the
model, although this difference needs to be confirmed by an analysis
in which the observed and predicted quantities are more obviously comparable.
One possible
way of strengthening the evolution in the simulation would
to reduce the strength of the feedback in massive galaxies (Katsianis
private communication).

\subsection{A new model - the flakey faucet model}

In this section we present a heuristic model 
that can reproduce the shape of the GS and the rapid
low-redshift evolution. It is based on a model
proposed by Peng et al. (2010) to explain the stellar
mass functions of red and blue galaxies which we
have modified to explain the new observations.

Peng et al. (2010) showed that the 
difference in the stellar mass functions of the red and blue
galaxies can be explained by a simple quenching model.
In their model a galaxy evolves along the GMS with its star-formation
rate being proportional to its stellar mass,
thus producing a horizontal GMS on a plot of SSFR
versus stellar mass,
until a catastrophic quenching event
occurs; the galaxy then moves rapidly to the `red and dead' region of the
diagram. The difference between the stellar mass functions
of red and blue galaxies (e.g. Baldry et al.
2012) can be explained almost exactly if
the probability of quenching is also proportional
to stellar mass. However, this model doesn't explain the curved
GS and the rapid low-redshift evolution.

The difference between our model and that of Peng et al.
is that we assume something milder occurs
when the galaxy is quenched. We assume that the quenching simply consists of
the gas supply to the galaxy being turned off. We
discuss possible physical explanations of this stochastic disruption
of the gas supply in \S 5.4. After the gas supply is turned off, we model
the evolution of the galaxy over the SSFR-versus-stellar-mass diagram;
it is this post-quenching evolution that produces the
curved GS and the rapid low-redshift evolution.

Many of the details of the models are the same.
We assume that as long as gas is
being supplied to a galaxy, its star-formation rate
is proportional to its stellar mass. We also assume like Peng et al. that the
mean SSFR of the galaxies to which gas is still being supplied
decreases with cosmic time.
Using a slight modification of
equation 1 of Peng et al., the SSFR of an
indvidual galaxy, as long as gas is being supplied to it, is given by:

\smallskip
\begin{align}
{SFR \over M_*} = SSFR = 2.5 \left({t \over 3.5}\right)^{-2.2}\ Gyr^{-1} + k
\end{align}
\smallskip

\noindent The constant $k$ represents the position of a galaxy relative to
the mean SSFR at that time. We assume $k$ remains a constant 
as long as gas is being
supplied to the galaxy.
We also assume, like Peng et al., that the quenching probability
is proportional to the galaxy's stellar mass:

\smallskip
\begin{align}
prob_{quenching} = c \times M_*
\end{align}
\smallskip
\noindent The constant $c$ is the first of the parameters of our model.

To model the evolution of the galaxy after the gas supply is turned
off, we need to know the gas mass at that time,
which is given by:

\smallskip
\begin{align}
M_{gas} = SFR / \epsilon
\end{align}
\smallskip
\noindent where $SFR$ is the star-formation rate
and $\epsilon$ is the star-formation efficiency, which is the second of the
two parameters in our model. With this equation and the following two
equations, we can follow a galaxy's evolution over
the SSFR-versus-stellar-mass diagram once the galaxy's gas supply
has been turned off:

\smallskip
\begin{align}
\Delta M_* = SFR \Delta t
\end{align}
\smallskip
\begin{align}
\Delta M_{gas} = -SFR \Delta t
\end{align}
\smallskip
\noindent in which $\Delta t$ is
an interval of cosmic time. 

There are some implicit assumptions behind these equations.
We are assuming, as we did in \S 4.4, that the star-formation rate is proportional to the
gas mass rather than to a different power of the gas mass (there is evidence
for both in the literature - Kennicutt and Evans 2012). Also, despite the
evidence in this paper (\S 4.4) and elsewhere that the star-formation
efficiency falls with decreasing SSFR, for simplicity we assume
it is constant. 

We created a realisation of this model by continuously
injecting stochastically galaxies onto
the diagram with a stellar mass of $10^8 M_{\odot}$ and
an SSFR given by equation 14, starting at
a redshift of 4.0 and continuing to the present time. 
In giving a value of $k$ to each galaxy, we assumed
that the distribution of $k$ is a
Gaussian
distribution with a standard deviation of
0.2 in $\rm log_{10}(SSFR)$. 
Once the galaxy is injected, we follow its motion accross the diagram, as long
as the gas supply is switched on, using equations 14 and 17. In each time
step, we use equation 15 and a random number generator
to determine whether to switch off the gas supply;
once the gas is switched off, we follow the evolution of the
galaxy 
using equations 16-18.

We assumed a star-formation efficiency ($\epsilon$ in equation 16)
of $\rm 10^{-9}\ year^{-1}$, a typical value for
optically-blue galaxies (\S 4.4, Figure 9).
The only other parameter in this model is the quenching probability ($c$ in equation
15), which
we adjusted until we got 
a galaxy distribution in the Universe today that 
looked like the observed GS.
We found that we got reasonable agreement
with the observations if  

\smallskip
\begin{align}
prob_{quenching}(M_*) = \left( {M_* \over 3.3 \times 10^{9}} \right)\ Gyr^{-1}
\end{align}
\smallskip

Figure 10 shows the distribution of the galaxies in a plot of SSFR
versus stellar mass 
at a redshift of 0 for one realization of this model. 
The dashed lines show the region
in the bottom left of the diagram where the HRS is incomplete
(Section 2). If one mentally excludes the galaxies
in Figure 10 that fall in this region, the figure's resemblance 
to Figure 1, the GS as seen by the HRS, is quite good.
The model explains the curvature of the GS 
as the result of the curved path an individual galaxy
follows in this diagram once its gas supply has been cut off.
We emphasise that the only thing we have done to get this
agreement between
the model and the observations is to adjust
one parameter:
the normalization in the quenching equation (equation 19).

The simplicity of the model means that it
is very easy to see what would happen if we change
some of its features.
In the model, there are no outflows once the gas
supply
has been turned off. If there continued to be outflows, galaxies would
evolve down the diagram more rapidly, reducing the density of points in
the figure.
On the other hand, if star-formation efficiency decreases with
decreasing SSFR, galaxies would move more slowly down the diagram,
leading to a greater number
of galaxies at intermediate values of SSFR.

Since Fig. 10 shows that most of the
galaxies with stellar masses $>10^{10}\ M_{\odot}$
in the Universe today are now evolving as closed boxes, the model 
will naturally
lead to strong low-redshift evolution.
Dunne et al. (2011) found that by a
redshift of 
$\simeq0.45$ galaxies contain roughly
five times more dust (and therefore gas) than today.
The simplicity of the model 
makes a quantitative comparison
with the observations pointless, but a rough calculation shows
that the model should give approximately the right amount of evolution.
If $\epsilon$ is the star-formation efficiency, the ratio
of the gas mass at time $t$ to the gas mass at $t=0$ is given
by $M_g/M_0 = e^{-\epsilon t}$. 
An increase of a factor of 5 in the gas mass by
$z = 0.45$  requires
a star-formation efficiency of $\rm \epsilon = 3.5 \times 10^{-10}\ year^{-1}$.
This is close to the centre of the histogram of star-formation
efficiencies in Figure 9.
As long as most galaxies are now evolving
as closed boxes, the observed evolution in gas mass is easy to
explain.

Although this model is outwardly not very different from
the model of Peng et al. (2010), its assumptions
about the
galaxy population are very different.
In our model, there
is no longer a star-forming GMS
and an area of `red-and-dead' galaxies. 
Galaxies are still forming stars all the way down the Galaxy Sequence.
The only physical distinction is between
galaxies which are still being supplied by gas (the horizontal
stub of galaxies in Figure 10) and the galaxies to which the gas supply
has been shut off but which are still forming stars.

\subsection{Where's the physics?}

The heuristic model described in the last section was successful in reproducing
in a qualitative, and at least in a semi-quantitative way some of the key
properties of the galaxy population.
What physical process could lie behind it? Any process must pass three tests. First,
it must be a `weak quenching' process, stopping any further supply of gas to the
galaxy but not removing the gas that is already in the
galaxy. Second, it must work in all
environments, since the shape of the GS is similar in the field and in clusters (Eales
et al. 2017; this paper; O17) and the rapid low-redshift evolution is also a feature
of the overall galaxy population not just the
galaxies in clusters\footnote{The Butcher-Oemler effect in clusters may have
been an early example of this rapid low-redshift evolution (Butcher and Oemler
1978).}. Third, it must have the stochastic element necessary to explain the
different stellar mass functions of early- and late-type galaxies.

Many processes suggested as quenching processes fail one or more of these tests.
Ram-pressure stripping of gas by an intracluster medium (Gunn and Gott 1972) fails
the first and second. Removal of the gas in a galaxy by wind or jet from an AGN
or starburst - `feedback' - fails the first. Note that we do not claim these processes are
not occurring - there is plenty of evidence for feedback, for example (e.g. Cicone et al. 2014) - merely
that they are not the processes responsible for switching off the future gas
supply to galaxies.
Galactic `strangulation' or `starvation' (Larson et al. 1980), 
in which the extended gaseous envelope around a galaxy
is stripped away, is the kind of weak quenching process we require, since once the envelope
is removed
there will be no further inflow of gas onto the galaxy, 
but it was orginally suggested as a process that would take place in dense environments,
thus failing test 2.

In the remainder of this section, we describe one process that potentially passes all
three tests, although at least part of this is speculation and
we do not claim there are no other possibilities.

The mass in the quenching equation (equation 19) is not a characteristic
mass because changing the time unit, from Gyr to years for example,
changes its value.
However, we can derive a characteristic mass using the
natural timescale for the growth of a galaxy, $SSFR^{-1}$.
Rewriting equation 19 as the probability of quenching in this
time period rather than per Gyr, the characteristic 
stellar mass is $\rm \simeq4 \times 10^8\ M_{\odot}$,
$\rm 3 \times 10^9\ M_{\odot}$ and $\rm 9 \times 10^9\ M_{\odot}$
at $z=0$, $z=1$ and $z=2$, respectively.
These correspond to masses of the surrounding dark-matter halos
of $\rm \simeq 6 \times 10^{10}\ M_{\odot}$, $\rm 4\times10^{11}\ M_{\odot}$
and $\rm 8 \times 10^{11}\ M_{\odot}$, respectively (Moster, Naab and White 2013).
Ideally, the physical process we require should pass both the three tests
above and also explain these characteristic masses.

Our speculative process was inspired by numerical simulations
of the way gas cools onto galaxies.
These simulations show that
gas accretes on to galaxies in two ways: a `hot mode' in which shocks
are set up in the gas, heating it to the virial temperature, from which it
slowly cools and is accreted by the galaxy; a `cold mode' in which gas falls freely
onto a galaxy along
cold flows (Katz et al. 2003; Keres et al. 2005, 2009; Nelson et al. 2013).
The simulations imply the 
transition halo mass below which gas can be supplied efficiently via cold flows
is $\rm 10^{11}-10^{11.5}\ M_{\odot}$, fairly similar to the values
we require. This process therefore provides a possible explanation
of the characterisitic mass that we observe, although it is worth noting that
(a) the transition from cold mode to hot mode may actually be rather gentle, occurring
over a range of halo mass (Nelson et al. 2013), and (b) our attempt to use
lensed {\it Herschel} sources to estimate the mass above which baryons cannot easily accrete on to
galaxies gives a value $\simeq$10 times higher than the
predictions of the numerical simulations (Amvrosiadis et al. 2017).

The other requirement is stochasticity. 
Dekel and Birnboim (2006) show that cold flows can still sometimes
penetrate to the central galaxy even in halos above the transition mass.
Since the number of cold flows in an individual halo is small,
we speculate that whether gas can be still supplied onto a galaxy
in a halo above the transition mass may depend
on the particular substructure in the individual halo, thus introducing
the element of stochasticity that we need.

\subsection{Other evidence for a unitary population} 

There is other recent evidence that instead of two separate classes
of galaxies, there is actually a unitary population. 

The first evidence comes from the ATLAS$^{3D}$ survey of 260
ETGs. The survey has shown
that 86\% of ETGs have the velocity field
expected for
a rotating disk (Emsellem et al. 2011); for
92\% of these `fast rotators'
there is also photometric evidence for a stellar disk (Krajnovic
et al. 2013).
Cappellari et al. (2013) used the ATLAS$^{3D}$ results
to propose that
there is a gradual change in galaxy properties from LTGs
to ETGs
rather than a dichotomy at the ETG/LTG boundary. The only
exception are
the 14\% of the ETGs that are `slow-rotators', for
which there is generally (but not always) no photometric evidence
for a stellar disk (Krajnovic et al. 2013).
Very recently, Cortese et al. (2016), using integral-field spectroscopy
from the SAMI survey, have shown that in terms of their kinematic properties
LTGs and fast-rotator ETGs form a continuous class of objects.

\begin{figure}
\centering
\includegraphics[width=74mm]{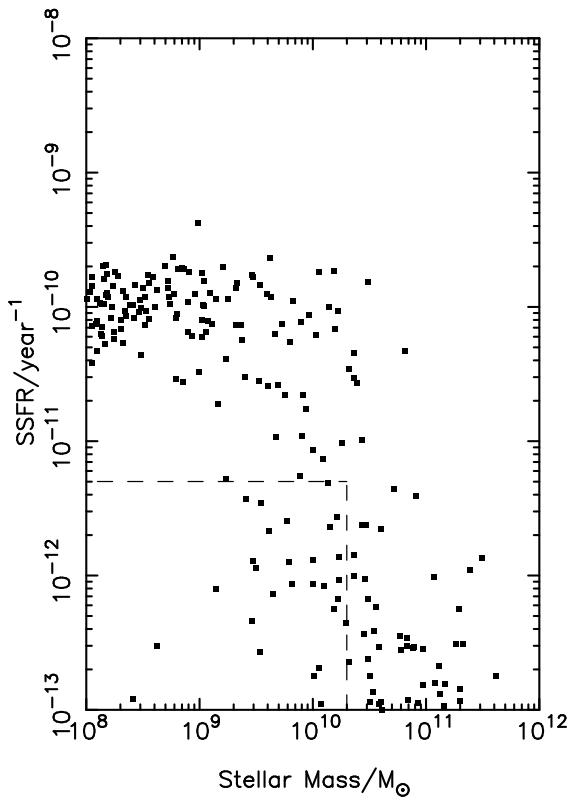}
  \caption{Plot of SSFR
against stellar mass at a redshift of 0 
for one realisation of the flaky faucet model.
The figure should be compared with the empirical version of the
GS shown in Figure 1. The galaxies in the box in the bottom
left-hand corner would not have been detected in the HRS, and
so galaxies in this box do not appear in Figure 1.
}
\end{figure}

Other evidence comes from surveys of the ISM
in ETGs. 
The best evidence for the existence of a cool ISM in many
ETGs comes from the {\it Herschel} observations of the
ETGs in the {\it Herschel} Reference Survey.
Smith et al. (2012b) detected dust emission from 50\% of the HRS ETGs;
{\it Herschel} observations of a much larger sample of ETGs drawn from
the ATLAS$^{3D}$ survey have detected a similar percentage
(Smith et al. in preparation).
The big increase
in detection rate from ground-based CO observations of ETGs (e.g.
Young et al. 2011) to {\it Herschel} observations
suggests that the common assumption that ETGs do not contain
a cool ISM is largely a function of instrumental sensitivity -
if we had more sensitive instruments we would find an even higher fraction
with a cool ISM.
Interferometric
observations
of the molecular gas 
in ETGs generally show a rotating disk similar to
what is seen in LTGs (Davis et al. 2013).

Thus these recent results 
are consistent with our argument that
there is a single population of galaxies (with the possible exception of the slow-rotating
ETGs - \S 5.6).
In our heuristic model (\S 5.3),
the dust and gas detected in ETGs is the
residual ISM left at the end of their evolution down the GS -
these galaxies are ones in which the gas supply was
shut off early and have been evolving as closed boxes ever
since.
A common counter-argument used against this idea is that 
the kinematics of the gas in ETGs are characteristic
of gas acquired as the result of recent
mergers.
However, the 
fact that very few
ETGs have a gas rotation axis pointing in the opposite direction to the
stellar rotation axis, and in the vast majority
of the cases the two rotation axes are broadly
in the same direction, is strong evidence that most of the
gas is this residual ISM
(see Eales et al. (2017) for this argument in more detail).

\subsection{Steps to a new paradigm}

The main argument we have tried to make in this paper and a previous
paper (Eales et al. 2017) is that the galaxyscale revealed by
{\it Herschel} does not look much like the predictions
of the current paradigm, in which
there are two distinct classes of galaxy and
a violent quenching process moves a galaxy from one class to the
other. The curved Galaxy Sequence shown in Figure 1, with the gradual
change in galaxy morphology along it, seems to us to require a gentler
quenching process. The one we have suggested in Section 5.3 is
the stochastic removal of the gas supply (a possible physical
process is suggested in \S 5.4). If we are correct,
the motion of galaxies along the Galaxy Sequence, as well as the rapid 
low-redshift evolution, is caused, not by the actual removal of the
gas supply,
but by the gradual evolution that occurs once the
gas is turned off.

A number of other groups have also recently investigated quenching
processes in the galaxy population. 
Cassado et al. (2015) found evidence for a weak quenching process
in low-density environments but for strong quenching in clusters,
whereas Schawinski et al. (2014) 
found evidence
for weak quenching in late-type galaxies but strong quenching in early-type
galaxies.
Peng, Maiolino and Cochrane (2015) used the metallicity
distributions of star-forming and passive galaxies to
argue that the evolution from one population to the other
must have occurred over $\simeq$4 billion years - evidence
for weak quenching.

O17 reached a similar conclusion to us about the
distribution of galaxies in the plot of SSFR versus 
stellar mass, but reached a slightly different conclusion about its
significance. They concluded that there are actually three types
of galaxy: star-forming galaxies, genuinely passive galaxies with no ISM,
and an intermediate class in which galaxies are still forming stars
but with a lower efficiency than in the first class (essentially
our optically-red class). 
The gradual change in morphology along the GS seen in Figure 1
seems to us evidence against this idea.
Like us, however,
O17 concluded that the
evolution from the star-forming class to the intermediate class
occurs as the result of the consumption of gas by star formation.
Very recently, Bremer et al. (in preparation)
have looked at the 
SEDs and structures of galaxies in 
the intermediate region of the SSFR-versus-stellar-mass plot, concluding like
us that the evolution accross this region is the result of the
gradual consumption of gas by the formation of stars in the disks
of galaxies.

Although there is accumulating evidence for this gentler version of galaxy evolution,
there
are some unanswered questions.

First, is there a subset of ETGs that are genuinely passive
galaxies
formed by some violent evolutionary process? One possible
class are the 14\% of ETGs that are slow rotators (\S 5.5). 
One way to test this possibility would be to
measure the star-formation rate and map the ISM
in {\it all} ETGs, in order to look for the
residual star formation and disks that should be
present if 
the only evolutionary process is the weak quenching
process we have proposed.
Unfortunately, the required observations get extremely challenging for
galaxies with $log_{10}(SSFR) < -11.5$.

The second bigger question, for which there is also not a definitive
answer in the current paradigm, is what causes the morphological variation
along the GS seen in Figure 1?
We don't have an answer but a possible clue is that the GS
is also a sequence of time: the epoch during which most of the galaxy's stars formed moves
earlier as we move down the GS.
One idea for forming a galaxy's bulge is that it formed
as
the result of the rapid motion of star-forming clumps towards
the galaxy's centre (Noguchi et al. 1999; Bournaud et al.
2007; Genzel et al. 2011, 2014).
If this process worked better at earlier times, this 
might explain the morphological change along the
GS.

\section{Summary}

The {\it Herschel} surveys
gave a radically different view of galaxy
evolution from optical surveys.
We first list our basic observational
results.

\begin{enumerate}

\item Rather than a star-forming Galaxy
Main Sequence (GMS) and a separate region of red-and-dead
galaxies, two 
{\it Herschel} surveys, selected in
completely different ways, 
reveal a single, curved Galaxy Sequence (GS)
extending
to low
values of specific star-formation rate (Figures 1 and 3).
The curvature of the GS confirms other recent results
(Whitaker et al. 2014; Gavazzi et al. 2015;
Lee et al. 2015; Schreiber
et al. 2016; Tomczak et al. 2016; O17) and we show in
\S 5.5 that there is plenty of evidence from other wavebands
for this picture of a unitary galaxy population rather than
two populations.

\item The
star-formation efficiency (star-formation rate divided
by gas mass) falls as one moves down the GS (\S 4.4).

\item Galaxy morphology gradually changes as one moves down the 
GS rather
than there being a jump from early-type to late-type
galaxies (Fig. 1)

\item 15-30\% of the galaxies in the far-infrared
{\it Herschel}
ATLAS would have been classified
as passive galaxies based on their optical colours. We use stacked spectra
to show that these red colours are caused by an old stellar population
rather than reddening by dust. Nevertheless, 
these galaxies still contain significant reservoirs of interstellar gas, and the
stacked spectra confirm that
they are still forming stars. They are red but not dead.
After correcting for Malmquist bias, we find that the space-density
of optically-red star-forming galaxies, which are 
missed by optical studies,
is at least as high as the space-density of optically-blue
star-forming galaxies of the same stellar mass.

\item The 250-$\mu$m luminosity function of both optically-blue and
optically-red galaxies shows
rapid evolution at low redshift, with the evolution appearing stronger for
the red galaxies (\S 4.3). 

\item We use stacked optical spectra of the H-ATLAS
galaxies at $0.3 < z < 0.4$ to show that
the galaxies responsible for this rapid evolution
have a significant spectral break at 4000\ \AA\, explaining the red optical
colours,
but also bright emission lines and a $UV$ upturn,
implying
a high star-formation rate.
These galaxies are red but very lively.
It seems likely that these galaxies, which have high stellar
masses, high star-formation rates and,
even a few billion years in the past, an old stellar population, are the
relatively recent ancestors of the
ETGs in the Universe today.

\end{enumerate}

We have explored whether existing galaxy-evolution models
can explain these results. Our concluson that
galaxies are a unitary population lying on a single GS
removes the need for a rapid quenching process.
We show that the popular
gas regulator, or bathtub, model cannot explain 
the curved GS and the strong low-redshift
evolution, and the EAGLE numerical galaxy simulation does
not reproduce the strong low-redshift evolution. We propose an alternative model in
which the gas supply to galaxies is stochastically cut off (the flaky faucet
model). We show that this model provides a natural explanation of
the curved GS, while still explaining earlier results
such as the difference between the stellar mass functions of optically-red
and optically-blue galaxies. 
The model naturally explains the rapid low-redshift evolution
because most
massive galaxies are no longer being supplied by gas
and are now evolving as closed boxes.

\section*{Acknowledgments}

We are grateful to the many scientists who have contributed to
the success of the {\it Herschel} ATLAS and the {\it Herschel} Reference
Survey.
We thank Antonio Kitsianis for a useful e-mail interchange about his
paper on the EAGLE predictions and for supplying a digital version
of one of his figures.
We also thank an anonymous referee for comments that significantly improved the
paper, in particular its
readability outside the
far-infrared club. 
EV and SAE acknowledge funding
from the UK Science and Technology Facilities Council consolidated grant ST/K000926/1. 
MS and SAE have received
funding from the European Union Seventh Framework Programme  ([FP7/2007-2013]  
[FP7/2007-2011])  under  grant
agreement No. 607254.
PC, LD and SM acknowledge support from the European Research Council (ERC) 
in the form of Consolidator Grant {\sc CosmicDust} (ERC-2014-CoG-647939, PI H\,L\,Gomez). 
SJM LD and RJI acknowledge
support from the ERC in the form of the Advanced Investi-
gator Program,
COSMICISM
(ERC-2012-ADG
20120216,
PI  R.J.Ivison).  
GDZ acknowledges financial support from ASI/INAF agreement n.2014-024-R.0.
M.J.M.~acknowledges the support of the National Science Centre, Poland
through the POLONEZ grant 2015/19/P/ST9/04010;
this project has received funding from the European Union's Horizon
2020 research and innovation programme under the Marie
Sk{\l}odowska-Curie grant agreement No. 665778.

\bibliographystyle{mnras}

\begin{thebibliography}{99}
\bibitem[\protect\citeauthoryear{Abdo et al.}{2010}]{a10} Abdo, A. et al.
2010, 710, 133
\bibitem[\protect\citeauthoryear{Agius et al.}{2013}]{a13} Agius, N. et al. 2013,
MNRAS, 431, 1929
\bibitem[\protect\citeauthoryear{Amvrosiadis et al.}{2017}]{a17} Amvrosiadis, A.
et al. 2017, MNRAS, submitted
\bibitem[\protect\citeauthoryear{Baldry et al.}{2012}]{b12} Baldry, I.K.
et al. (2012), MNRAS, 441, 2440
\bibitem[\protect\citeauthoryear{Bauer et al.}{2013}]{b13} Bauer, A. et al.
2013, MNRAS, 434, 209
\bibitem[\protect\citeauthoryear{Bell et al.}{2004}]{b04} Bell, E.F. et al.
2004, ApJ, 608, 752
\bibitem[\protect\citeauthoryear{Bell and de Jong}{2001}]{b01} Bell, E.F. and
de Jong, R.S. 2001, ApJ, 550, 212
\bibitem[\protect\citeauthoryear{Bendo et al.}{2010}]{b10} Bendo, G.J. et al.
2010, MNRAS, 402, 1409
\bibitem[\protect\citeauthoryear{Blanton et al.}{2003}]{b2003} Blanton, M.R. et
al. (2003), AJ, 125, 2348
\bibitem[\protect\citeauthoryear{Blanton and Roweis}{2007}]{br7} Blanton, M.R. and
Roweis, S. 2007, AJ, 133, 734
\bibitem[\protect\citeauthoryear{Boselli and Gavazzi}{2006}]{b06} Boselli, A.
\& Gavazz, G. 2006, PASP, 118, 517
\bibitem[\protect\citeauthoryear{Boselli et al.}{2010}]{b2010} Boselli, A.
et al. 2010, PASP, 122, 261
\bibitem[\protect\citeauthoryear{Bournaud et al.}{2007}]{b2007}
Bournaud, F., Elmgreen, B.G. \& Elmgreen, D.M.
2007, ApJ, 670, 237
\bibitem[\protect\citeauthoryear{Bouch\'e et al.}{2010}]{bou10} Bouch\'e, N. et al.
2010, ApJ, 718, 1001
\bibitem[\protect\citeauthoryear{Bourne et al.}{2012}]{bou1} Bourne, N. et al. 2012,
MNRAS, 421, 3027
\bibitem[\protect\citeauthoryear{Bourne et al.}{2016}]{b16} Bourne,
N. et al. 2016, MNRAS, 462, 1714
\bibitem[\protect\citeauthoryear{Bruzual and Charlot}{2003}]{b2003} Bruzual,
G. \& Charlot, S. 2003, MNRAS, 344, 1000
\bibitem[\protect\citeauthoryear{Butcher and Oemler}{1978}]{b1978} Butcher,
H. \& Oemler, A. 1978, ApJ, 219, 18
\bibitem[\protect\citeauthoryear{Cappellari et al.}{2013}]{c2013} Cappellari,
M. et al. 2013, MNRAS, 432, 1862
\bibitem[\protect\citeauthoryear{Casado et al.}{2015}]{c2015} Casado, J., Acasibar,
Y., Gavilan, M., Terlevich, R., Terlevich, E., Hoyos, C. \& Diaz, A.I.
2015, MNRAS, 451, 888
\bibitem[\protect\citeauthoryear{Chabrier}{2003}]{c2003} Chabrier, G. 2003,
PASP, 115, 763
\bibitem[\protect\citeauthoryear{Charlot and Fall}{2000}]{c2000} Charlot,
S. \& Fall, S.M. 2000, ApJ, 539, 718
\bibitem[\protect\citeauthoryear{Cicone et al.}{2014}]{c2014} Cicone, C. et
al. 2014, A\&A, 562, 21
\bibitem[\protect\citeauthoryear{Ciesla et al.}{2012}]{ci2012} Ciesla, L.
et al. 2012, A\&A, 543, 161
\bibitem[\protect\citeauthoryear{Cortese et al.}{2012}]{c2012} Cortese,
L. et al. 2012, A\&A, 540, 52
\bibitem[\protect\citeauthoryear{Cortese et al.}{2014}]{c2014} Cortese, L.
et al. 2014, MNRAS, 440, 942
\bibitem[\protect\citeauthoryear{Cortese et al.}{2016}]{c2016} Cortese, L.
et al. 2016, MNRAS, 463, 170
\bibitem[\protect\citeauthoryear{Da Cunha et al.}{2008}]{dc2008} Da Cunha, E.,
Charlot, S. and Elbaz, D. 2008, MNRAS, 388, 1595
\bibitem[\protect\citeauthoryear{Daddi et al.}{2007}]{d2007} Daddi, E. et al.
2007, ApJ, 670, 156
\bibitem[\protect\citeauthoryear{Dariush et al.}{2011}]{d2011} Dariush, A.
et al. 2011, MNRAS, 418, 64
\bibitem[\protect\citeauthoryear{Dariush et al.}{2016}]{d2016} Dariush, A.
et al. 2016, MNRAS, 456, 2221
\bibitem[\protect\citeauthoryear{Davies et al.}{2016}]{d16a} Davies, L.J.M. et al.
2016, MNRAS, 461, 458
\bibitem[\protect\citeauthoryear{Davis et al.}{2013}]{d13} Davis, T.A. et al. 2013,
MNRAS, 429, 534
\bibitem[\protect\citeauthoryear{De Vis et al.}{2017}]{d2017} De Vis, P.
et al. 2017, MNRAS, 464, 4680
\bibitem[\protect\citeauthoryear{Dekel and Birnboim}{2006}]{d06} Dekel, A. and
Birnboim, Y. 2006, MNRAS, 368, 2
\bibitem[\protect\citeauthoryear{Draine et al.}{2007}]{d07} Draine, B.T. et
al. 2007, ApJ, 663, 866
\bibitem[\protect\citeauthoryear{Driver et al.}{2009}]{d09} Driver, S. et al.
2009, Astron. Geophys., 50, 12
\bibitem[\protect\citeauthoryear{Driver et al.}{2016}]{d16} Driver, S. et al.
2012, ApJ, 827, 108
\bibitem[\protect\citeauthoryear{Dunne et al.}{2011}]{d2011} Dunne, L. et al.
2011, MNRAS, 417, 1510
\bibitem[\protect\citeauthoryear{Dye et al.}{2010}]{dy1} Dye, S. et al. 2010,
AA, 518, L10
\bibitem[\protect\citeauthoryear{Eales et al.}{2009}]{e2009} Eales, S. et al. 2009,
ApJ, 707, 1779
\bibitem[\protect\citeauthoryear{Eales et al.}{2010}]{e0} Eales, S. et al. 2010, PASP,
122, 499
\bibitem[\protect\citeauthoryear{Eales et al.}{2012}]{e1} Eales, S. et al. 2012,
ApJ, 761, 168
\bibitem[\protect\citeauthoryear{Eales et al.}{2015}]{e2015} Eales, S.A. et al.
2015, MNRAS, 452, 3489
\bibitem[\protect\citeauthoryear{Eales et al.}{2017}]{e2017} Eales, S.A. et al. 2017,
MNRAS, 465, 3125
\bibitem[\protect\citeauthoryear{Edge et al.}{2013}]{e2013} Edge, A., Sutherland, W.,
Kuijken, K., Driver. S., McMahon, R., Eales, S. \& Emerson, J.P. 2013, The
Messenger, 154, 32
\bibitem[\protect\citeauthoryear{Elbaz et al.}{2007}]{e2007} Elbaz, D.
et al. 2007, A\&A, 468, 33
\bibitem[\protect\citeauthoryear{Emsellem et al.}{2011}]{e2011} Emsellem,
E. et al. 2011, MNRAS, 414, 888
\bibitem[\protect\citeauthoryear{Faucher-Gigu\`ere et al.}{2011}]{fg1} Faucher-Gigu\`ere,
C.A., Keres, D. \& Ma, , C.-P. 2011, MNRAS, 417, 2982
\bibitem[\protect\citeauthoryear{Gavazzi et al.}{2015}]{g15} Gavazzi, G. et
al. 2015, A\&A, 580, 116
\bibitem[\protect\citeauthoryear{Genzel et al.}{2011}]{g2011} Genzel, R.
et al. 2011, ApJ, 733, 30
\bibitem[\protect\citeauthoryear{Genzel et al.}{2014}]{g2014} Genzel, R.
et al. 2014, ApJ, 785, 75
\bibitem[\protect\citeauthoryear{Genzel et al.}{2015}]{g2015} Genzel, R.
et al. 2015, ApJ, 800, 20
\bibitem[\protect\citeauthoryear{Groves et al.}{2015}]{g2015} Groves, B. et al.
2015, ApJ, 799, 96
\bibitem[\protect\citeauthoryear{Gunn and Gott}{1972}]{g1972} Gunn, J.E. \& Gott, J.R.
1972, ApJ, 176, 1
\bibitem[\protect\citeauthoryear{Hardcastle et al.}{2016}]{h16} Hardcastle,
M. et al. 2016, MNRAS, 462, 1910
\bibitem[\protect\citeauthoryear{Hayward and Smith}{2015}]{h15} Hayward, C. and Smith,
D.J.B. 2015, MNRAS, 446, 1512
\bibitem[\protect\citeauthoryear{Hopkins et al.}{2013}]{h13} Hopkins, A. et al. 2013,
MNRAS, 430, 2047
\bibitem[\protect\citeauthoryear{Hughes et al.}{2013}]{h13} Hughes, T.M., Cortese, L., Boselli,
A., Gavazzi, G. \& Davies, J.I. 2013, A\&A, 550, 115
\bibitem[\protect\citeauthoryear{Ilbert et al.}{2013}]{i13} Ilbert, O. et al.
2013, A\&A, 556, 545
\bibitem[\protect\citeauthoryear{James et al.}{2002}]{ja1} James, A., Dunne, L., Eales, S.
and Edmunds, M. 2002, MNRAS, 335, 753
\bibitem[\protect\citeauthoryear{Katsianis et al.}{2017}]{kat2017} Katsianis,
A. et al. 2017, MNRAS, in press (arXiv: 1708.01913)
\bibitem[\protect\citeauthoryear{Katz et al.}{2003}]{k2003} Katz, N., Keres, D., 
Dav\'e, R. \& Weinberg, D.H. 2003
in J.L. Rosenberg \& M.E. Putman ed., The IGM/Galaxy
Connection. The Distribution of Baryons at $z=0$,
Vol. 281 of Astrophysics and Space Science Library,
How Do Galaxies Get Their Gas? p185
\bibitem[\protect\citeauthoryear{Kennicutt}{1998}]{k1998} Kennicutt, R.C.
1998, ARAA, 36, 189
\bibitem[\protect\citeauthoryear{Kennicutt and Evans}{2012}]{k2012} Kennicutt, R.C.
\& Evans, N.J. 2012, ARAA, 50, 531
\bibitem[\protect\citeauthoryear{Keres et al.}{2005}]{k2005} Keres,
D., Katz, N., Weinberg, D.H. \& Dav\'e, R. 2005, MNRAS,
363, 2
\bibitem[\protect\citeauthoryear{Keres et al.}{2009}]{k2009} Keres, N.
\& Hernquist, L. 2009, ApJ, 700, L1
\bibitem[\protect\citeauthoryear{Krajnovic et al.}{2013}]{k2013} Krajnovic,
D. et al. 2013, MNRAS, 432, 1768
\bibitem[\protect\citeauthoryear{Larson et al.}{1980}]{l1980} Larson, R.B., Tinsley,
B.M. \& Caldwell, C.N. 1980, ApJ 237, 692
\bibitem[\protect\citeauthoryear{Lee et al.}{2015}]{L2015} Lee, N. et al.
2015, ApJ, 801, 80
\bibitem[\protect\citeauthoryear{Lilly et al.}{2013}]{li1} Lilly, S.J., Carollo, M.C., Pipino,
A., Renzini, A. \& Peng, Y. 2013, ApJ, 772, 119
\bibitem[\protect\citeauthoryear{Liske et al.}{2015}]{L15} Liske, J. et al.
2015, MNRAS, 452, 2087
\bibitem[\protect\citeauthoryear{Loveday et al.}{2012}]{L12} Loveday, J. et al.
(2012), MNRAS, 420, 1239
\bibitem[\protect\citeauthoryear{Marchetti et al.}{2016}]{m16} Marchetti,
L. et al. 2016, MNRAS, 456, 1999
\bibitem[\protect\citeauthoryear{Martig et al.}{2009}]{m2009} Martig, M.,
Bournaud, F., Teyssier, R. \& Dekel, A. 2009, ApJ, 707, 250
\bibitem[\protect\citeauthoryear{Moster, Naab and White}{2013}]{m2013} Moster, B.J., Naab,
T. and White, S.D.M. 2013, MNRAS, 428, 3121
\bibitem[\protect\citeauthoryear{Nelson et al.}{2013}]{n2013} Nelson, D., Vogelsberger, M.,
Genel, S., Sijacki, D., Keres, D., Springel, V. \& Hernquist, L. 2013, MNRAS, 429, 3353
\bibitem[\protect\citeauthoryear{Noguchi}{1999}]{n1999} Noguchi, M.
1999, ApJ, 514, 77
\bibitem[\protect\citeauthoryear{Noeske et al.}{2007}]{n2007} Noeske, K.G. et al.
2007, ApJ, 660, L43
\bibitem[\protect\citeauthoryear{Oemler et al.}{2017}]{o2017} Oemler, A., Abramson,
L.E., Gladders, M.D., Dressler, A., Poggianti, B.M. \& Vulcani, B. 2017,
ApJ, 844, 45 (O17)
\bibitem[\protect\citeauthoryear{Page and Carrera}{2000}]{pc2000} Page,
M. \& Carerra, F. 2000, MNRAS, 311, 433
\bibitem[\protect\citeauthoryear{Papadopoulos and Geach}{2012}]{pg12} Papadopoulos,
P. and Geach, J.E. 2012, ApJ, 757, 157
\bibitem[\protect\citeauthoryear{Peng et al.}{2010}]{p2010} Peng,
Y.-J. et al. 2010, ApJ, 721, 193
\bibitem[\protect\citeauthoryear{Peng and Maiolino}{2014}]{pe2} Peng, Y.-J. and Maiolino, R.
2014, MNRAS, 443, 3643
\bibitem[\protect\citeauthoryear{Peng et al.}{2015}]{p2015} Peng, Y., Maiolino, R.
\& Cochrane, R. 2015, Nature, 521, 192
\bibitem[\protect\citeauthoryear{Pilbratt et al.}{2010}]{p2010} Pilbratt, G. et
al. 2010, A\&A, 518, L1
\bibitem[\protect\citeauthoryear{Pineda et al.}{2013}]{p2013} Pineda, J.L. et al.
2013, A\&A, 554, 103
\bibitem[\protect\citeauthoryear{Planck Collaboration}{2011}]{p2011} Planck Collaboration
2011, A\&A, 536, 19
\bibitem[\protect\citeauthoryear{Planck Collaboration}{2014}]{p2014} Planck Collaboration
2014, A\&A, 571, 26
\bibitem[\protect\citeauthoryear{R\'emy-Ruyer et al.}{2014}]{r2014} R\'emy-Ruyer, A.
et al. 2014, A\&A, 563, 31
\bibitem[\protect\citeauthoryear{Renzini and Peng}{2015}]{r2015} Renzini, A.
\& Peng, Y. 2015, ApJ, 801, L29
\bibitem[\protect\citeauthoryear{Rodighiero et al.}{2011}]{r2011} Rodighiero, G. et
al. 2011, ApJ, 739, 40
\bibitem[\protect\citeauthoryear{Rowlands et al.}{2012}]{r2012} Rowlands, K. et al.
2012, MNRAS, 419, 2545
\bibitem[\protect\citeauthoryear{Rowlands  et al.}{2014}]{r2014} Rowlands, K.
et al. 2014, MNRAS, 441, 1017
\bibitem[\protect\citeauthoryear{Saintonge et al.}{2012}]{sa2} Saintonge, A. et al. 2012,
758, 73
\bibitem[\protect\citeauthoryear{Sandstrom et al.}{2013}]{sa13} Sandstrom, K. et al.
2013, ApJ, 777, 5
\bibitem[\protect\citeauthoryear{Santini et al.}{2014}]{s2014} Santini, P.
et al. 2014, A\&A, 562, 30
\bibitem[\protect\citeauthoryear{Schawinski et al.}{2014}]{sc2014} Schawinski,
K. et al. 2014, MNRAS, 440, 889
\bibitem[\protect\citeauthoryear{Schreiber et al.}{2016}]{sc2016} Schreiber,
C. et al. 2016, A\&A, 589, 35
\bibitem[\protect\citeauthoryear{Scoville et al.}{2014}]{sc1} Scoville, N.Z. et al. 2014,
2014, ApJ, 783, 84
\bibitem[\protect\citeauthoryear{Scoville et al.}{2016}]{s2016} Scoville, N.
et al. 2016, ApJ, 820, 83
\bibitem[\protect\citeauthoryear{Speagle et al.}{2014}]{s2014} Speagle, J.S.,
Steinhardt, C.L., Capak, P.L. \& Silverman, J.D. 2014, ApJS, 214, 15
\bibitem[\protect\citeauthoryear{Smith et al.}{2011}]{s1} Smith, D.J.B. et al. 2011, MNRAS,
416, 857
\bibitem[\protect\citeauthoryear{Smith et al.}{2012a}]{s2} Smith, D.J.B. et al. 2012a, MNRAS,
427, 703
\bibitem[\protect\citeauthoryear{Smith et al.}{2012b}]{s3} Smith, M.W.L. et al.
2012b, ApJ, 748, 123
\bibitem[\protect\citeauthoryear{Smith et al.}{2012c}]{sm1} Smith, M.W.L. et al.
2012c, ApJ, 756, 40
\bibitem[\protect\citeauthoryear{Smith et al.}{2016}]{s4} Smith, M.W.L. et al.
in preparation
\bibitem[\protect\citeauthoryear{Tacconi et al.}{2010}]{t2010} Tacconi, L.
et al. 2010, Nature, 463, 781
\bibitem[\protect\citeauthoryear{Taylor et al.}{2011}]{ta1} Taylor, E. et al. 2011, MNRAS,
418, 1587
\bibitem[\protect\citeauthoryear{Thomas et al.}{2005}]{t05} Thomas, D., Maraston, C., Bender, R. and
Mendes de Oliveira, C. 2005, ApJ, 621, 673
\bibitem[\protect\citeauthoryear{Toomre}{1977}]{t1977} Toomre, A.
1977, {\it Evolution of Galaxies and Stellar Populations, Proceedings
of a Conference at Yale University}, edited
by B.M. Tinsley and R.B. Larson (Yale
University Observatory, New Haven, Conn.), 401
\bibitem[\protect\citeauthoryear{Tomczak et al.}{2016}]{t16} Tomczak, A.
et al. 2016, ApJ, 817, 118
\bibitem[\protect\citeauthoryear{Tremonti et al.}{2004}]{tr} Tremonti, C. et al. 2004,
ApJ, 613, 898
\bibitem[\protect\citeauthoryear{Wang et al.}{2016}]{w16} Wang, L. et al. 2016,
MNRAS, submitted
\bibitem[\protect\citeauthoryear{Wang et al.}{2016}]{w16a} Wang, L. et al. 2016a,
A\&A, 592, L5
\bibitem[\protect\citeauthoryear{Wang et al.}{2016}]{w16b} Wang, L. et al. 2016b,
MNRAS, 461, 1719
\bibitem[\protect\citeauthoryear{Whitaker et al.}{2012}]{w12} Whitaker, K. et al.
2012, ApJ, 754, 29
\bibitem[\protect\citeauthoryear{Whitaker et al.}{2014}]{w14} Whitaker, K. et al.
2014, ApJ, 795, 104
\bibitem[\protect\citeauthoryear{Valiante et al.}{2016}]{v16} Valiante, E.
et al. 2016, MNRAS, 562, 3146
\bibitem[\protect\citeauthoryear{Young et al.}{2011}]{y2011} Young, L.M. et al.
2011, MNRAS, 414, 940
\bibitem[\protect\citeauthoryear{Zibetti et al.}{2009}]{z09} Zibetti, S.,
Charlot, S. \& Rix, H.-W. 2009, MNRAS, 400, 1181

\end{thebibliography}

\bsp

\label{lastpage}

\end{document}